# Targeted learning: Towards a future informed by real-world evidence

Susan Gruber[1], Rachael V. Phillips[2], Hana Lee[3], Martin Ho[4], John Concato[5], Mark J. van der Laan[2]

[1]Putnam Data Sciences, LLC, Cambridge, Massachusetts, United States
[2]Department of Biostatistics, School of Public Health, University of California at Berkeley, Berkeley, California, United States
[3]Office of Biostatistics, Center for Drug Evaluation and Research, United States Food and Drug Administration, Silver Spring, Maryland, United States
[4]Biostatistics, Google, Mountain View, California, United States
[5]Office of Medical Policy, Center for Drug Evaluation and Research, United States Food and Drug Administration, Silver Spring, Maryland, United States

**Abstract:** The 21st Century Cures Act of 2016 includes a provision for the U.S. Food and Drug Administration (FDA) to evaluate the potential use of real-world evidence (RWE) to support new indications for use for previously approved drugs, and to satisfy post-approval study requirements. Extracting reliable evidence from real-world data (RWD) is often complicated by a lack of treatment randomization, potential intercurrent events, and informative loss to follow up. Targeted Learning (TL) is a sub-field of statistics that provides a rigorous framework to help address these challenges. The TL Roadmap offers a step-by-step guide to generating valid evidence and assessing its reliability. Following these steps produces an extensive amount of information for assessing whether the study provides reliable scientific evidence in support regulatory decision making. This paper presents two case studies that illustrate the utility of following the roadmap. We use targeted minimum loss-based estimation combined with super learning to estimate causal effects. We also compared these findings with those obtained from an unadjusted analysis, propensity score matching, and inverse probability weighting. Non-parametric sensitivity analyses illuminate how departures from (untestable) causal assumptions would affect point estimates and confidence interval bounds that would impact the substantive conclusion drawn from the study. TL's thorough approach to learning from data provides transparency, allowing trust in RWE to be earned whenever it is warranted.

**Acknowledgement**

The authors gratefully acknowledge funding from the United States Food and Drug Administration (US FDA) pursuant to Contract 75F40119C10155. The content is the view of the author(s), and does not necessarily represent the official views of, nor an endorsement by, the FDA/HHS or the U.S. Government.

# 1. Introduction

The 21st Century Cures Act of 2016 included a provision for the U.S. Food and Drug Administration (FDA) to evaluate the potential use of real-world evidence (RWE) to support new indications for use for previously approved drugs, and to satisfy post-approval study requirements (FDA 2020). The FDA's draft framework (2018) describes the real-world data (RWD) sources that can generate RWE, and outlines challenges and opportunities in developing rigorous approaches to producing reliable evidence (FDA 2018).

RWD capture patient health status and health care deliver from a variety of sources, including electronic health records (EHR), medical claims, product and disease registries, patient-generated data, and wearable devices. Analyses of RWD produce RWE representing clinical evidence regarding the usage and potential risks and benefits of medical products. In addition to traditional randomized controlled trials (RCT), other study designs—including pragmatic trials, externally controlled trials, and observational studies—increasingly rely on RWD to produce RWE. Despite legitimate concerns regarding the completeness and accuracy of information on covariates, ascertainment of exposures and outcomes, as well as lack of randomization and loss to follow-up (LTFU), RWD can be a source of valuable insights (Corrigan-Curay et al. 2018, Simon et al 2021). Key considerations for regulatory decision-making include whether the RWD are fit for purpose, whether the study provides adequate scientific evidence, and whether study conduct meets regulatory requirements (FDA 2018, FDA 2021). This paper reports on an FDA-funded project to explore the use of Targeted Learning (TL) as a principled approach to incorporating RWD into regulatory decision making.

TL is a statistical framework for efficient learning from data (van der Laan and Rose 2011). The TL estimation roadmap offers step-by-step guidance for transforming a precise statistical question that meets the substantive goal of the study into reliable RWE. The roadmap addresses all components of the ICH E9(R1) Guideline definition of a statistical estimand: population, treatment, outcome variable, summary measure, and intercurrent events (ICH 2020, Gruber, et al. 2021). Within the TL framework, targeted minimum loss-based estimation (TMLE) and super learning (SL) are the recommended statistical methodologies for providing efficient estimation and valid inference (van der Laan and Rose 2011). TMLE can appropriately adjust for sources of bias, including from intercurrent events occurring after treatment initiation, that impact the outcome. From a causal perspective, an intercurrent event can alter the outcome or induce potentially informative missingness that can bias causal effect estimates. TMLE can adjust for this bias when sufficient confounder information is available.

SL is a vital tool for mitigating model misspecification bias. In high dimensional settings, e.g., where rich medical histories have been documented in EHR, unbiased effect estimation in finite datasets often requires some form of dimension reduction. With SL, this task can be done sensibly using a variety of data adaptive techniques. Recent applications of SL include risk score prediction (Pirrachio et al. 2015, Gruber et al. 2020), identifying health outcomes of interest (Carrell et al. 2022), and estimating causal effects of treatments and exposures in observational and randomized studies (Balzer et al. 2019, Kreif et al. 2017, Kempker et al. 2020).

RCT findings are considered the benchmark research design because causal assumptions (described below) are presumed to be met. However, intercurrent events such as treatment switching, LTFU, and other forms of right censoring introduce non-randomized elements into RCTs that bias unadjusted effect estimates. We demonstrate that TL can be successfully applied to RCTs whether or not these problems arise. A recent publication estimates that 99% of RCTs fail to report on missing data separately by outcome, and 95% fail to report a method for judging risk of bias associated with missing data (Kahale et al. 2019). When missingness is potentially informative, unbiased estimation requires analyzing the RCT data as if it were observational. It is vitally important that appropriate methods are well understood, and their use becomes mainstream.

This paper demonstrates how to integrate TL into regulatory decision making. We first review the TL Estimation Roadmap, then illustrate its utility in two case studies. In each we evaluated the marginal additive treatment effect (ATE), using data made publicly available by study authors. Analyses of these data, and a closely related simulation study, illustrate the performance of TMLE+SL compared with the unadjusted estimator, and two popular propensity score (PS) based methodologies—PS matching (Rosenbaum and Rubin 1983), and inverse probability weighting (IPW) (Hernán et al. 2000).

## 2. The Targeted Learning Estimation Roadmap

The TL estimation roadmap offers a step-by-step guide to estimating causal effects from data (Petersen and van der Laan 2014, Ho et al. 2021). The process begins with a statistical question satisfying the substantive goals of the study that can be answered from data, and a precise description of the experiment giving rise to the data. Steps 1-3 of the TL roadmap define the statistical and causal models, causal parameter of interest, statistical estimand, and identifying assumptions that link what can be estimated from data with the causal question of interest. This approach naturally integrates intercurrent events and the other ICH-defined elements of a causal estimand. Step 4 of the roadmap involves estimating the statistical parameter and its uncertainty. Step 5 is to conduct sensitivity analyses to assess the level of support in the data for a causal interpretation of the study finding and substantive conclusion.

Step 1. The statistical model is a collection of possible probability distributions of the data consistent with the data generating experiment that gave rise to the data over time, including intercurrent events. The statistical model should be defined broadly enough to ensure that the true distribution is not excluded. Incorporating domain knowledge, such as known bounds on the outcome, is helpful, but the model should not be further restricted based on assumptions not definitively known to be true.

Step 2. The causal parameter, $\psi^{causal}$, is a feature of the true data distribution, defined in terms of a causal model of the full (counterfactual) data, rather than as a coefficient in a particular parametric model. The causal model expresses conditional independencies, assumes exogenous errors, and should be in alignment with the statistical model.

Step 3. The statistical estimand, $\psi^{obs}$, is a quantity that can be estimated from observed data that best approximates $\psi^{causal}$. $\psi^{obs}$ and $\psi^{causal}$ are linked through identifying assumptions informally known as 1) *positivity*, an assumption that within strata defined by confounders there is a non-zero probability of receiving treatment at all levels of interest; 2) *consistency*, an assumption that the outcome observed under the assigned treatment is equivalent to the counterfactual outcome under that level of exposure; and 3) *coarsening at random* (CAR), which implies no unmeasured confounding. PS diagnostics (e.g., summaries of covariate balance, distribution of the PS in each treatment arm and/or IP weights, C-statistic) can indicate whether the positivity assumption is met, but the other two assumptions are not testable from data. When all three assumptions hold, $\psi^{obs}$ is an expression of the causal quantity in terms of observable rather than counterfactual data.

Step 4. TMLE+SL is recommended for estimation of $\psi^{obs}$, due to its robustness finite sample performance and flexibility in addressing the challenges inherent in analyses of RWD. SL provides data adaptive modeling that mitigates model misspecification bias. TMLE respects bounds on the problem, and unlike PS-based methodologies, utilizes all available information in the data. This typically results in estimates with smaller mean squared error (MSE) than PS-based methods (Porter et al. 2011, Lendle et al. 2013). Because the sampling distribution of the TMLE is well understood, one can obtain valid inference, i.e., estimates of uncertainty, p-values, and 95% confidence intervals (CI), with good coverage properties.

Step 5. Sensitivity analyses help assess the robustness of conclusions drawn from the study findings. These should include a non-parametric investigation of the impact on results of any *causal gap*, the non-random portion of the difference between the true quantity of interest and its estimated value due to departures from the identifying assumptions (Díaz and van der Laan 2013). Domain knowledge, and an understanding of the experiment that gave rise to the data, can provide insight into the magnitude of realistic departures from the underlying assumptions. Understanding how point estimates and CI are affected is indicative of the level of support in the data for the substantive conclusion, and the reliability of the RWE.

## 3. Estimators of the ATE

Consider data consisting of $n$ independent and identically distributed (i.i.d.) observations, $O = (\Delta Y, \Delta, A, W)$, where $\Delta Y$ equals the outcome value when $\Delta = 1$, and is missing when $\Delta = 0$, $\Delta$ is a binary indicator of outcome missingness, $A$ is a binary treatment indicator, and $W$ is a vector of baseline covariates. The likelihood of the data can be factorized as $\mathcal{L}(O) = P(Y \mid \Delta, A, W)$, $P(\Delta \mid A, W), P(A \mid W), P(W)$. The TMLE literature refers to $P(Y \mid \Delta, A, W)$ and $P(W)$ as the $Q$ portion of the likelihood, and to $P(\Delta \mid A, W), P(A \mid W)$ as the $G$ portion of the likelihood (van der Laan and Rose 2011). The ATE parameter is defined in terms of the $Q$ portion of the likelihood, $\psi_{ATE}^{causal} = E(Y_1) - E(Y_0)$, where $Y_a$ is the counterfactual conditional mean outcome under exposure at level $A = a$, and the expectation is with respect to the distribution of $W$. When identifying assumptions are met, the corresponding statistical estimand in observed data is given by $\psi_{ATE}^{obs} = E(\Delta Y \mid \Delta = 1, A = 1, W) - E(\Delta Y \mid \Delta = 1, A = 0, W)$.

We estimated $\psi_{ATE}^{obs}$ using an unadjusted estimator, several variants of full PS matching, inverse probability of treatment weighting (IPTW) (or inverse probability of treatment and censoring weighting (IPTCW), when the data contained missing outcomes), and TMLE+SL. Each estimator's efficiency depends on the information in the data. The unadjusted estimator of the treatment effect uses a subset of information available in the $Q$ portion of the likelihood to estimate $\psi_{ATE}^{obs}$. PS matching uses information in only the PS component of $G$, $P(A\,|\,W)$. IPW estimators use information in both components of $G$. TMLE incorporates information from all components of $Q$ and $G$. Using more information in the data decreases asymptotic variance. Implementation details for each estimator are described next. All analyses were run in the R statistical programming environment, v3.6.1 (R Core Team, 2020).

### 3.1 PS Matching

Although PS matching is often used to estimate an average treatment effect among the treated, full matching can also be used to evaluate the ATE (Hansen 2004). Full matching on the logit of the PS was carried out using the *MatchIt* package, v3.0.2 (Ho et al., 2011). *MatchIt* produces weights proportional to each observation's contribution to one or more matched sets. These weights were incorporated into an unadjusted logistic regression of the outcome on treatment (variant 1), and as recommended in practice (Stuart 2010), an adjusted regression of the outcome on treatment and all covariates in the dataset (variant 2). Post-match adjustment is also recommended practice (Stuart 2010). The ATE estimates were evaluated by calculating the mean difference in predicted probabilities of experiencing the outcome when treated versus not treated.

For each matching variant, the PS was estimated using a main terms logistic regression model. As matches only consider the PS, ignoring the outcome, observations with missing outcomes were retained as potential matches, but then dropped from the subsequent analysis. Although common practice, and we use such PS matching estimators for comparison purposes, this approach should not be construed as the ideal version of matching. More recently proposed matching estimators, e.g., those that incorporate multiple imputation or that use machine learning to fine-tune the propensity score estimates to reduce covariate imbalances (Sekhon 2011, Imai and Ratkovic 2013, Ling et al. 2019), were not considered.

### 3.2 Inverse Probability Weighting

IPW estimates of the ATE were obtained from a weighted regression of the outcome, $Y$, on the binary treatment indicator, $A$. This univariate regression will yield the identical value for the ATE, though not the model coefficients, regardless of whether logistic or linear regression is used. We used linear regression, because the ingredients for evaluating a robust standard error for the ATE estimate (the coefficient in front of treatment in the model) were readily available. Robust SEs were calculated using the *sandwich* package v2.5-1 (Zeileis 2004, Zeileis et al. 2020).

Stabilized IP weights have been recommended over their unstabilized counterparts for reducing finite sample bias and variance (Hernán 2000). These weights were calculated as:

$$wt_{stab} = mean(\Delta) \frac{\Delta}{P(\Delta = 1|A, W)} \left[ mean(A) \frac{A}{PS} + mean(1-A) \frac{1-A}{1-PS} \right]. \quad (1)$$

When no observations are missing, this general formula reduces to the familiar IPTW equation for stabilized weights inside the square brackets. The IPW estimator is undefined when any term in the denominator of (1) equals zero, i.e., the positivity assumption is violated. Otherwise, an observation could receive an infinite IP weight. Recommended practice is to place on upper bound on the maximum allowed weight, to avoid overly large contributions from only a few influential observations (Cole and Hernán 2008). We investigated four truncation strategies, truncating the weights at either 40, 100, 1,000,000 (essentially, no truncation), and our recently proposed sample size-based bound of $\sqrt{n}\ln(n)/5$ (Gruber et al. 2022). For each strategy, any weight that exceeds the specified value is set to the allowed maximum.

The PS and missingness probabilities were estimated using a main terms logistic regression model regressing treatment on all available covariates (variant 1) or using SL (variant 2).

### 3.3 TMLE+ SL

TMLE analyses were carried out using the *tmle* R package, v1.5.0-1 (Gruber and van der Laan 2012), and the *SuperLearner* R package, v2.0-24 (Polley et al. 2019). Except where stated, we used each package's default specifications. In particular, the SL library for modeling the outcome regression included linear regression, Bayesian additive regression trees (BART, in *dbarts* v0.9-11) (Chipman et al. 2010), and lasso (*glmnet* v3.0-2) (Friedman et al. 2010). The library for modeling the PS and missingness mechanisms ($G$ components) included logistic regression, BART, and generalized additive models (*gam* v1.16) (Hastie 2019).

Analogous to IPW, TMLE requires the product of the $G$ components of the likelihood to be bounded away from zero. We evaluated performance under four different PS truncation levels, equal to the reciprocal of the bounds on the IP weights: 0.025, 0.01, 0.000001, $5/[\sqrt{n}\ln(n)]$.

## 4. Case Study 1: A traditional randomized controlled trial with minimal loss to follow-up

Case study 1 illustrates how to apply each step of the roadmap and establishes baseline performance characteristics of each estimator in the context of a nearly ideal RCT (e.g., large number of randomized patients with minimal or non-informative LTFU and treatment non-adherence). An additional plasmode simulation study (Franklin et al. 2014) provides a comparison of performance under simulated selection bias and informative LTFU.

The International Stroke Trial (IST) was a multi-national RCT conducted in 1991-1996 to test the safety of aspirin and heparin in patients hospitalized for stroke (International Stroke Trial Collaborative Group 1997). Patients suspected of suffering an ischemic stroke ($n = 19{,}435$) were randomized to aspirin or no aspirin, with and without heparin, to assess several outcomes of interest. In an intent to treat (ITT) analysis the authors found a statistically significant effect of treatment with aspirin vs. no aspirin on recurrent ischemic stroke within 14 days after randomization ($\psi_0^{ATE} = -0.011, p < 0.001$). The authors found a non-statistically significant effect of treatment with aspirin vs. no aspirin on mortality within 14 days after randomization ($\psi_0^{ATE} = -0.004$, p > 0.05). De-identified data were downloaded from an online repository (Sandercock et al. 2011).

### 4.1 Complete case data analysis
The complete case analytic dataset contained $n = 19{,}408$ observations (dropping 27 observations where either outcome was missing). The data structure when all outcomes are observed simplifies to $O = (Y, A, W)$, where $W$ consists of 24 continuous, binary, and categorical covariates (Appendix Table A1).

*4.1.1 Applying the TL Roadmap*

Step 1. *Statistical model:* All distributions of the data consistent with study inclusion/exclusion criteria, respecting the process that gave rise to the data.

Step 2. *Causal model and causal parameter*: The causal parameter of interest reported in the original published study is the ATE, $\psi_{ATE}^{causal} = E(Y_1) - E(Y_0)$. Our causal model assumes exogenous errors, and independencies implied by the time ordering.

Step 3. *Statistical estimand and identifiability:* The statistical estimand is given by $\psi_{ATE}^{obs} = E(Y|A=1, W) - E(Y|A=0, W)$. Knowledge of the process that gave rise to the data aids in assessing the plausibility of the identifying causal assumptions. 1) The positivity assumption is guaranteed by baseline randomization, and by the nearly complete follow-up. 2) Adherence to assigned treatment was 88% or more in each treatment arm, with cessation largely due to hospital discharge. Under a slightly modified definition of treatment, treat within 48 hours of stroke, and continue until hospital discharge, 14-day adherence is essentially 100%. Thus, the consistency assumption is satisfied. 3) Baseline randomization, adherence to treatment, and complete follow-up imply the CAR assumption is satisfied.

Step 4. *Estimation:* PS matching, IPTW, and TMLE+SL were used to estimate the ATE. For PS matching we enforced an exact match on region by stratifying the matching procedure. For TMLE, SL was used for modeling the outcome, but we specified a parametric logistic regression model for the PS instead of using SL, because when analyzing RCT data theory guarantees consistency of the TMLE when PS estimates are modeled parametrically. The covariates incorporated as main terms in the PS model for the TMLE were pre-screened using a lasso algorithm that retains covariates associated with the outcome. Note that the automated screening process is an internal part of the TMLE estimation procedure, and thus adheres to the outcome blinding principle (for the analyst).

Step 5. *Sensitivity analyses and substantive conclusion:* All three identifying assumptions appear to be met. Randomization ensures the positivity assumption, and CAR assumptions are met as stated in Step 3. The consistency assumption is guaranteed in an ITT analysis. The causal gap is presumably quite close to zero. Nevertheless, results of a sensitivity analysis are presented in subsection 4.1.3 for illustration.

*4.1.2 Results*

*PS Diagnostics.* Common diagnostics allow us to understand how closely the treatment and control groups resemble each other. These diagnostics include a plot of the PS distribution within each treatment group showing near perfect overlap (Figure 1), the distribution of IP weights, and the standardized mean difference for all covariates before and after matching and IP weighting (Appendix Table A2). Given that randomization was successful, there were no imbalances before or after matching and IP weighting. The C-statistic of 0.51 indicates that the covariates are not predictive of treatment assignment. Stabilized IP weights were between 0.78 and 1.35 for all observations, with a mean of 1.0. Truncation at the levels we considered had no effect on the weights.

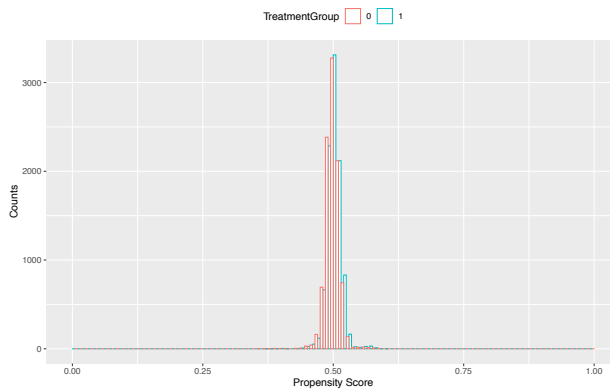

Figure 1. Subtask 1.1, Propensity score distribution in treated and control groups in the IST trial.

*Estimator Performance.* Estimates, standard errors (SE), 95% CIs, and p-values based on analytic estimates of the variance, as well as bootstrapped variance estimates for unadjusted, TMLE, and IPW estimates, are reported for the 14-day mortality outcome and the 14-day recurrent stroke outcome (Table 1). Because no IP weights or PS values were modified by truncation at any of the levels considered, only one IPW result and one TMLE result are displayed in the table.

Table 1. Point estimates, standard errors (SE), p-values, 95% confidence interval lower bounds (lb) and upper bounds (ub) based on analytic variance estimates and bootstrapped variance estimates for analyses of 14-day mortality and 14-day recurrent ischemic stroke outcomes.

| Estimator | Estimate | Based on Analytic Variance | | | | Based on Bootstrap Variance | | | |
|---|---|---|---|---|---|---|---|---|---|
| | | SE | p-value | lb | ub | SE | p-value | lb | ub |
| *14-DAY MORTALITY[a]* | | | | | | | | | |
| **Unadjusted** | -0.004 | 0.004 | 0.381 | -0.011 | 0.004 | 0.004 | 0.391 | -0.012 | 0.005 |
| **Matching** unadj | -0.004 | -- | -- | -- | -- | 0.005 | 0.369 | -0.013 | 0.005 |
| adj | -0.003 | -- | -- | -- | -- | 0.004 | 0.518 | -0.012 | 0.006 |
| **IPTW** | -0.003 | 0.004 | 0.407 | -0.011 | 0.005 | 0.004 | 0.388 | -0.011 | 0.004 |
| **TMLE** | -0.003 | 0.004 | 0.575 | -0.011 | 0.006 | 0.004 | 0.530 | -0.010 | 0.005 |
| | | | | | | | | | |
| *14-DAY RECURRENT ISCHEMIC STROKE[b]* | | | | | | | | | |
| **Unadjusted** | -0.009 | 0.002 | < 0.001 | -0.013 | -0.005 | 0.002 | < 0.001 | -0.013 | -0.005 |
| **Matching** unadj | -0.009 | -- | -- | -- | -- | 0.002 | < 0.001 | -0.014 | -0.004 |
| adj | -0.009 | -- | -- | -- | -- | 0.002 | < 0.001 | -0.013 | -0.004 |
| **IPTW** | -0.009 | 0.002 | < 0.001 | -0.013 | -0.005 | 0.002 | < 0.001 | -0.013 | -0.005 |
| **TMLE** | -0.009 | 0.002 | < 0.001 | -0.013 | -0.005 | 0.002 | < 0.001 | -0.013 | -0.005 |

[a]Reported ATE for mortality = -0.004; [b]Reported ATE for stroke outcome = -0.011.

All causal methodologies produced results quite similar to the unadjusted estimate, and to published findings. Treatment has essentially no impact on 14-day mortality but confers slight protection against 14-day recurrent stroke. Minor differences from published results may stem from the fact that our dataset contained slightly fewer observations due to missing outcomes. Bootstrap estimates of the SEs are in close agreement with analytic SE estimates (not available for the post-match adjusted matching estimator). All estimators have similar SEs. Because the outcome is rare, there is little variation in the outcome to explain, so little opportunity for variance reduction.

*4.1.3 Sensitivity analysis and substantive conclusion*

Since all three identifying assumptions are met by the study design and conduct, in our analysis the only possible source of any causal gap is that our complete case analysis omitted 27 of the 19,435 observations in the data. The most extreme bias would be observed if either all omitted observations in the treatment arm were true cases and all omitted observations in the control arm were non-cases, or the reverse. We therefore created these two hypothetical versions of the data and carried out an unadjusted regression of the outcome on treatment. The point estimates, SE, and 95% CIs indicate there is essentially no change in the point estimate and 95% CI at the most extreme plausible values for the causal gap (Table 2).

Table 2. Sensitivity analysis results from unadjusted regression of the 14-day mortality and stroke outcomes on treatment in the original complete case analysis, an analysis in which all missing outcomes are set to 0 for treated subjects and 1 for control subjects, and an analysis where all missing outcomes are set to 1 for treated subjects and 0 for control subjects.

| Imputed value for missing outcomes | 14-day Mortality | | 95%CI | | 14-day Stroke | | 95%CI | |
| --- | --- | --- | --- | --- | --- | --- | --- | --- |
| | Est | SE | lb, | ub | Est | SE | lb, | ub |
| complete case | -0.0036 | 0.0040 | -0.0115 | 0.0044 | -0.0092 | 0.0020 | -0.0132 | -0.0052 |
| treated=0, control = 1 | -0.0049 | 0.0041 | -0.0128 | 0.0031 | -0.0104 | 0.0021 | -0.0144 | -0.0064 |
| treated=1, control = 0 | -0.0023 | 0.0041 | -0.0102 | 0.0056 | -0.0076 | 0.0021 | -0.0116 | -0.0036 |

As Figure 2 illustrates, with respect to 14-day mortality (left), the lower bound on the 95% CI is negative under most of the values of the causal gap investigated ("Adj Units" are the ratio of the causal gap to the amount of adjustment due to measured covariates). Thus, this analysis strongly supports the conclusion that treatment does not increase risk of mortality. Because the CI around the actual study finding includes the null, and there is little reason to think the causal gap is large and also supports the conclusion that treatment has no impact on mortality. However, the analysis provides essentially no support for a conclusion that treatment increases risk of mortality, because for that inference to be true the causal gap would have to be 3.4 times as large as the adjustment due to measured confounders.

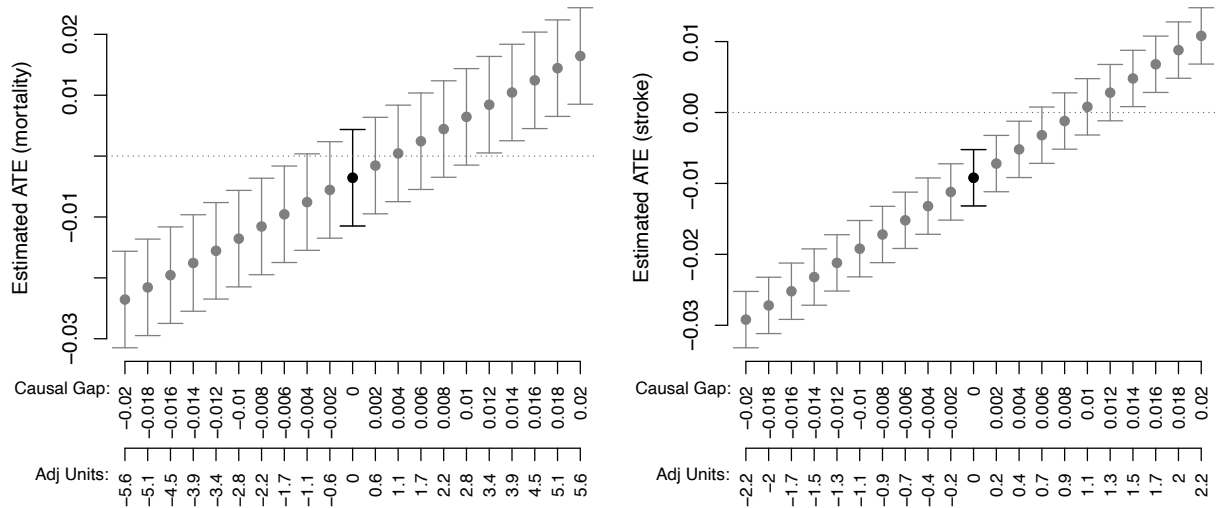

Figure 2.Case study 1: Sensitivity analysis showing the effect on point estimates and 95% confidence interval bounds under presumed causal gap up to +/-0.02 for 14-day mortality (l) and 14-day recurrent stroke (r). The "Adj Units" axis shows the ratio of the causal gap to the amount of adjustment due to measured variables. Study finding shown in black.

With respect to 14-day recurrent stroke (Figure 2, right), the analysis clearly supports the conclusion that treatment decreases risk for stroke. The causal gap would have to be nearly 70% larger than the effect size for the 95% CI to include the null. Recall that adjusting for potential baseline confounders had no impact on point estimates, so a causal gap of that magnitude in this well designed and well conducted study is practically impossible. Taken together, the sensitivity analyses demonstrate strong support for the substantive conclusions.

*4.2 Plasmode simulation studies*

Plasmode simulation studies were designed to investigate estimator performance when informative selection into treatment and informative LTFU introduce bias into the unadjusted causal effect estimate. For each study we analyzed 1000 datasets by drawing observations with replacement from the original IST data, generating an artificial PS, treatment indicator ($A$), outcome ($Y$), and missingness indicator ($\Delta$), conditional on the covariates; 5%, 15%, and 25% informative missingness was imposed in studies I, II, and III, respectively.

The PS was generated as a complex function of the only six confounders in the dataset, SEX, AGE, RCONSC_F, RSBP, RDELAY, RVISINF_N (see Appendix Table A3 for covariate definitions). We defined $Z_1 = -2 + 2.5\text{SEX} + \text{RCONSC\_F} - 4.5\text{SEX} \times \text{RCONSC\_F}$, $Z_2 = \text{AGE/RSBP}$, $Z_3 = \log(\text{RDELAY}/10)$, and $Z_4 = \log(\text{AGE})$, and then calculate the PS as $P(A = 1|W) = expit(-1.8 - 0.5 RVISINF\_N + 1.1 Z_1 Z_2 - Z_3 + 0.4 Z_4)$. The coefficients in the PS model were chosen so that approximately 15% of observations were in the treatment group, and to introduce confounding bias.

A binary outcome, $Y$, was generated conditional on treatment, $A$, and a subset of covariates $W'$ consisting of 26 main term covariates and the same four transformed versions of baseline covariates, $Z_1, Z_2, Z_3, Z_4$, $P(Y = 1|A, W) = expit(\beta_0 + \beta_1 A + \beta_2 AZ_1 + \boldsymbol{\beta}^T \boldsymbol{W'})$. In order to make the relationships in the simulated data similar to those in the original dataset, the values for the 30 coefficients, $\boldsymbol{\beta}$, were set equal to the estimated coefficients in a logistic regression of the true mortality outcome on the main term covariates in $\boldsymbol{W'}$ (Appendix Table A3). Then $\beta_0, \beta_1$, and $\beta_2$ were chosen so that the outcome occurred in approximately 17% of observations, and the treatment effect was heterogeneous. For these data the true ATE is given by $\psi^{ATE} = -0.1381$.

After each dataset was generated, we generated missingness indicators for studies I, II, and III as follows,
I. $P(\Delta_I = 1|A, W') = expit(0.9 + 1.15(0.8 STYPE\_LACS + 0.3 RDEF5\_N + 0.34 Z_4 + 0.2A)$,
II. $P(\Delta_{II} = 1|A, W') = expit(-0.065 + 0.8 STYPE\_LACS + 0.3\ RDEF5\_N + 0.34 Z_4 + 0.2A)$,
III. $P(\Delta_{III} = 1|A, W') = expit(-0.85 + 1.07(0.8 STYPE\_LACS + 0.3 RDEF5\_N + 0.34 Z_4 + 0.2A))$,

and then set $Y_{study,i}$ to missing when $\Delta_{study,i} = 0$.

*4.2.1 Applying the TL Roadmap*

Definitions of the statistical model, causal model, target causal parameter, and statistical estimand defined in Steps 1-3 of the roadmap, remain unchanged. However, the Step 3 assessment of the identifying assumptions linking the observed data to the full data need to be re-examined in light of selection bias and LTFU.

*4.2.2 Results*

*PS Diagnostics.* These are derived from a single dataset consisting of covariates from the original observations, and simulated treatment. The plot of PS distributions within each treatment group shows that although there are some differences between the two groups, there is considerable overlap (Figure 3). The C-statistic of 0.78 supports this conclusion.

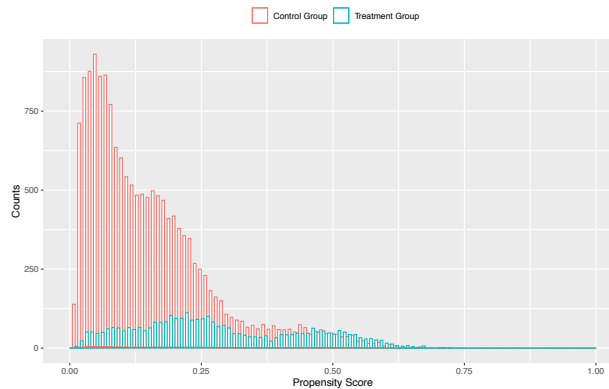

Figure 3. Simulation study: Distribution of propensity scores in plasmode dataset, in which 17% of observations are treated.

Standardized mean differences in the observed data indicate that covariate imbalances are greatly reduced after matching or IP weighting (Appendix Table A4). The IPTC weights account for imbalances due to both selection into treatment and missingness. The stabilized IPTC weights ranged between 0 and 16 in all studies, so truncation again had no impact on the results (Table 3). As expected, the mean stabilized weight is approximately equal to the proportion of observations where the outcome is observed. Although it seems counterintuitive that the maximum observed weight is largest when only 5% of observations are missing, and smallest when 25% are missing, this situation occurred because observations where subjects happened to receive rare treatment assignments tended to be missing more often. When the outcome is missing, the observation is assigned a weight of zero.

Table 3. Minimum, mean, 99th percentile, and maximum value of IPTC weights in simulation studies I, II, and III, as estimated by logistic regression (GLM) and super learner (SL).

| Study | Estimation method | min | mean | 99th percentile | max |
|---|---|---|---|---|---|
| I. (5% missing) | GLM | 0 | 0.96 | 2.89 | 15.82 |
| | SL | 0 | 0.94 | 2.78 | 10.42 |
| II. (15% missing) | GLM | 0 | 0.86 | 2.62 | 14.92 |
| | SL | 0 | 0.84 | 2.61 | 9.04 |
| III. (25% missing) | GLM | 0 | 0.75 | 2.47 | 13.64 |
| | SL | 0 | 0.74 | 2.48 | 10.13 |

Table 4. Mean bias, standard deviation of the bootstrap estimates (SE), MSE, and coverage of 95% CI based on the bootstrap SE and the estimated analytic SE for simulation studies I, II, and III.

| Estimator | | Bias | SE | MSE | 95% CI Coverage Bootstrap SE | Analytic SE |
|---|---|---|---|---|---|---|
| *I. 5% Missing* | | | | | | |
| Unadjusted | | 0.0722 | 0.0065 | 0.00526 | 0 | 0 |
| Matching | unadj | -0.0327 | 0.0101 | 0.00117 | 0.091 | 0.109 |
| | adj | 0.0136 | 0.0068 | 0.00023 | 0.476 | -- |
| IPTCW | GLM | 0.0047 | 0.0055 | 0.00005 | 0.862 | 0.906 |
| | SL | 0.0021 | 0.0050 | 0.00003 | 0.927 | 0.962 |
| TMLE | | 0.0012 | 0.0050 | 0.00003 | 0.948 | 0.950 |
| *II. 15% Missing* | | | | | | |
| Unadjusted | | 0.0736 | 0.0067 | 0.00546 | 0 | 0 |
| Matching | unadj | -0.0302 | 0.0110 | 0.00104 | 0.217 | 0.228 |
| | adj | 0.0153 | 0.0074 | 0.00029 | 0.458 | -- |
| IPTCW | GLM | 0.0046 | 0.0058 | 0.00006 | 0.878 | 0.915 |
| | SL | 0.0021 | 0.0053 | 0.00003 | 0.934 | 0.962 |
| TMLE | | 0.0011 | 0.0053 | 0.00003 | 0.943 | 0.944 |
| *III. 25% Missing* | | | | | | |
| Unadjusted | | 0.0755 | 0.0071 | 0.00574 | 0 | 0 |
| Matching | unadj | -0.0271 | 0.0117 | 0.00087 | 0.359 | 0.378 |
| | adj | 0.0177 | 0.0077 | 0.00037 | 0.353 | -- |
| IPTCW | GLM | 0.0048 | 0.0063 | 0.00006 | 0.872 | 0.924 |
| | SL | 0.0022 | 0.0058 | 0.00004 | 0.934 | 0.960 |
| TMLE | | 0.0014 | 0.0056 | 0.00003 | 0.950 | 0.954 |

*Estimator Performance.* Mean bias, standard deviation of the bootstrap estimates (SE), MSE, as well as coverage of 95% CI based on the bootstrap SE and the estimated analytic SE for all three simulation studies, are shown in Table 4. PS Matching was less biased than the unadjusted estimate, but model misspecification and inappropriate handling of missing outcomes produced biased estimates with large SEs. Post-match adjustment greatly reduced the MSE, and somewhat improved the CI coverage. IPTCW that relied on parametric modeling was fairly successful, but incorporating SL greatly reduced bias, and slightly reduced the SE. This more favorable ratio of bias to SE manifests as greatly improved CI coverage (Gruber et al. 2022). In all scenarios TMLE had the smallest bias and MSEs, and nearly optimal CI coverage.

## 5 Case Study 2: Pragmatic trial with 25% loss to follow up

The Acupuncture for Chronic Headache in Primary Care trial (ACHPC) was a pragmatic trial carried out in England and Wales in 1999-2001 to assess the effect of acupuncture in practice on headache (Vickers et al. 2004). De-identified data were made publicly available by the study team (Vickers 2006). Adult general care patients having chronic headache (n=401) were randomized to either 3 months of acupuncture (12 treatments), or usual care. Randomization was site-specific, conditional on age, sex, headache type, baseline headache score, and number of years of headache disorder. Randomization probabilities were calculated for each subject to correct for chance imbalances among subjects who were previously randomized (Vickers et al. 2004). The target statistical estimand was the intention-to-treat (ITT) effect of acupuncture vs. no acupuncture on headache score, an ATE. Unbiased causal effect estimation was complicated by LTFU. The primary outcome measure, mean weakly headache score (MWHS) at the 12-month end of study, was missing in 25% of observations. Study authors reported results from a complete case analysis that adjusting for baseline determinants of randomization probabilities, and from sensitivity analyses where multiple imputation was used to generate missing outcome values. In a separate publication, the authors demonstrated a clear effect of treatment randomization on a correlated outcome, a global headache metric, that was available for 95% of subjects (Vickers and McCarney 2003). We focus on the original MWHS outcome, to illustrate how the TL roadmap informs causal inference when some outcomes are missing, and to compare estimator performance in this setting.

### 5.1 Estimating the ATE: Point treatment analysis

The dataset consisted of $n = 401$ i.i.d. observations, $O = (\Delta Y, \Delta, A, W)$. Outcomes were missing for 100 (25%) of observations. These publicly available data provide an opportunity to apply methodologies that were not available at the time of the original analysis (TMLE+SL and IPW), and to evaluate the relative performance of TMLE, IPW, and PS Matching when there is substantial LTFU.

Patients who dropped out were slightly younger, had higher baseline headache scores, and were more likely to be in the control arm (22% treated vs. 29% control). Three covariates had missing values (pain at baseline, physical functioning, and role limitation physical at baseline, role limitation emotional at baseline). Prior to analyzing the data, missing covariate values were imputed by assigning the mode for binary variables, and the median for continuous variables. Corresponding binary missingness indicators were created added to the dataset.

*5.1.1 Applying the TL Roadmap*

Step 1. *Statistical model:* All distributions of the data consistent with study inclusion/exclusion criteria, respecting the process that gave rise to the data, e.g., treatment assignment was randomly assigned conditional on baseline covariates.

Step 2. *Causal model and causal parameter:* The causal parameter of interest reported in the original published study is the ATE, $\psi_{ATE}^{causal} = E(Y_1) - E(Y_0)$. For the ITT analysis the treatment of interest is defined as randomization to acupuncture ($A = 1$) vs no acupuncture ($A = 0$), under no LTFU. The causal model assumes exogenous errors, and conditional independencies implied by the time ordering.

Step 3. *Statistical estimand and identifiability:* The statistical estimand is given by $\psi_{ATE}^{obs} = E(\Delta Y | \Delta = 1, A = 1, W) - E(\Delta Y | \Delta = 1, A = 0, W)$. We note that this definition of the statistical estimand assumes, without proof, that intercurrent events are ignorable. In the next subsection we re-state the problem from a longitudinal perspective to avoid making that assumption. Assessment of Identifiability: 1) The positivity assumption is partially satisfied by randomization. However, if LTFU depleted a large proportion of observations within strata of confounders differentially by trial arm, there may be a practical violation of this assumption. PS diagnostics will provide helpful information for assessing whether this assumption is met. 2) For this ITT analysis the consistency assumption is easily satisfied, given that adherence to assigned treatment is immaterial. 3) The CAR assumption is satisfied with respect to treatment assignment by baseline randomization. However, it could be violated if there are unmeasured confounders of the associations between treatment, LTFU, and the outcome.

Step 4. *Estimation:* For all estimators, including TMLE, the PS was estimated using a main terms logistic regression model, conditioning on the five determinants of randomization probabilities, and an additional 10 variables selected using Lasso, based on their association with the outcome. Of note, such variables were identified by an automated process to ensure the outcome blinding principle. For IPTW the missingness mechanism was estimated once using logistic regression, and a second time using SL.
As stated in Section 3.1, observations were matched with and without allowing observations outside the area of common support to be discarded. After matching, observations where the outcome was missing were dropped from the analysis. ATE estimates were evaluated with and without post-match adjustment.
IPW and TMLE were extended to incorporate missingness probabilities, adjusting for the same 15 potential confounders. The IPTCW-GLM estimator incorporated estimates of the missingness probabilities based on logistic regression, a generalized linear model (GLM). The IPTCW-SL estimator incorporated estimates of the missingness probabilities based on SL. For TMLE, SL was also used to model the outcome regression and the missingness probabilities. All calls to SL specified 20-fold cross validation.

Step 5. *Sensitivity analyses and substantive conclusion*: Potential violations of the identifying causal assumptions discussed in Step 3 cast doubt on interpreting the findings as the true causal effect. Furthermore, if LTFU at different time points was driven by post-treatment covariates, there may be residual bias after adjusting for baseline covariates alone. The definition of the statistical estimand in Step 3 assumes without proof that intercurrent events are ignorable. A model-free, non-parametric sensitivity analysis will investigate how robust the conclusion is under potential causal gaps.

*5.1.2 Point treatment analysis results*

*PS Diagnostics.* The PS distributions in each treatment group, C-statistic of 0.62, and SMD tables all indicate the two groups are quite comparable at baseline (Figure 4, Appendix Table A3). IPTCW weights using GLM ranged from 0 to 4.20, with a mean of 0.74. When SL was used to estimate the missingness probabilities, the weights ranged from 0 to 3.28, with a mean of 0.71 (Appendix Table A4). There were no extreme weights, so truncation had no impact.

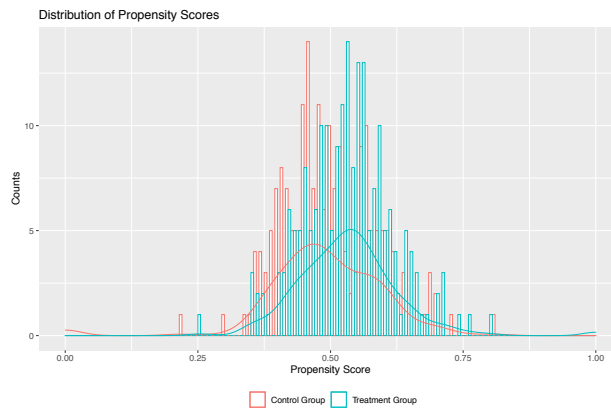

Figure 4. Propensity score distributions in treated and control groups in the ACHPC trial.

*Estimator Performance.* The estimated treatment effects, SEs, 95% CIs, and p-values are shown in Table 5. The bias cannot be calculated given that the true effect size is unknown, but PS-matching variants that discarded observations outside the areas of common support yielded estimates substantially closer to the null. Post-match adjustment moved these estimates away from the null and were in better agreement with IPTCW and TMLE. Using SL instead of logistic regression (GLM) for estimating the missingness probabilities moved IPTCW estimates further away from the unadjusted estimate, towards the null. Three estimators produced similar point estimates (PS matching, retain all observations, post-match adjustment, IPTCW-SL, TMLE), with TMLE having the smallest SE, and narrowest CI.

Table 5. Case study 2 estimates (Est) of the ATE in the entire dataset, standard error (SE), 95% confidence interval lower bound (lb) and upper bound (ub), and p-value.

| Estimator | Notes | Est | SE | 95% CI lb | 95% CI ub | p-value |
|---|---|---|---|---|---|---|
| **Unadjusted** | complete case | -6.10 | 1.77 | -9.57 | -2.62 | 0.0006 |
| **Matching (retain all observations)** | no post-match adjustment | -4.72 | 1.72 | -8.09 | -1.34 | 0.0062 |
| | post-match adjustment | -5.40 | 1.23 | -7.80 | -3.00 | <0.0001 |
| **Matching (discard observations)** | no post-match adjustment | -4.24 | 1.74 | -7.66 | -0.82 | 0.0150 |
| | post-match adjustment | -5.07 | 1.25 | -7.52 | -2.62 | 0.0001 |
| **IPTCW** | GLM for missingness | -5.68 | 1.43 | -8.48 | -2.88 | 0.0033 |
| | SL for missingness | -5.31 | 1.36 | -7.97 | -2.66 | 0.0036 |
| **TMLE** | | -5.31 | 1.20 | -7.67 | -2.95 | <0.0001 |

*5.1.3 Sensitivity analysis and substantive conclusion*

Sensitivity analyses reported in the original paper relied on parametric multiple imputation modeling assumptions and did not address bias due to potential unmeasured confounders. Nevertheless, they were intentionally designed to provide a conservative estimate of the treatment effect. Imputing values for the 100 missing outcomes yielded an estimated ATE of -3.91, consistent with the paper's substantive conclusion that acupuncture reduced MWHS.

We ran a non-parametric sensitivity analysis to evaluate the impact of an unknown causal gap on the substantive conclusion. PS diagnostics suggest the positivity assumption is met, and consistency is guaranteed in an ITT analysis. The only likely contributor to any causal gap is a violation of the CAR assumption with respect to time-varying confounding. The plot allows us to examine how large the residual bias would have to be to change the substantive conclusion (Figure 5). The x-axis is shown in units of the size of the causal gap, and in units of the ratio of the presumed causal gap to the bias adjustment due to measured confounders (0.079, the difference between the unadjusted and TMLE estimates of the ATE). The plot clearly shows that the causal gap would have to be at least 60% as large as the adjustment in the point estimate due to measured confounders to overturn the conclusion that acupuncture decreases 12-month MWHS. Concluding that acupuncture increased risk for MWHS would require much stronger residual bias. Clinicians and other experts with domain knowledge are in the best position to judge the plausible amount of the causal gap. The plot provides an understanding of the strength of support for the substantive conclusion at that level.

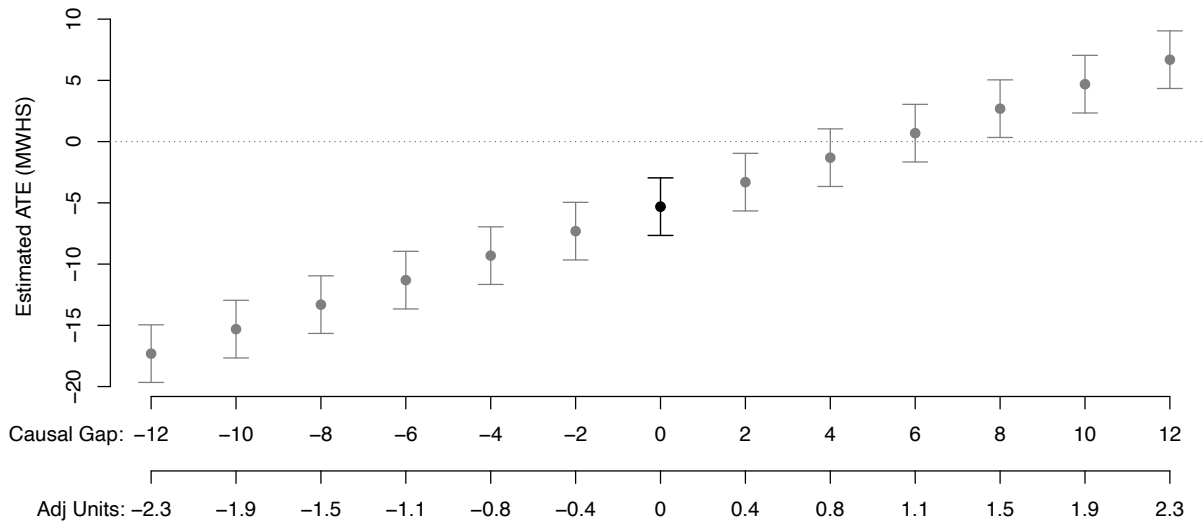

Figure 5. Case study 2: Sensitivity analysis showing the effect on point estimates and 95% confidence interval bounds under a presumed causal gap of up to +/-12. The "Adj Units" axis shows the ratio of the causal gap to the amount of adjustment due to measured variables. Study finding shown in black.

### *5.2 Estimating the ATE: Longitudinal Data Analysis*

Study authors collected data on medications, headache score, and care usage periodically throughout the study. We can exploit the availability of these time-varying covariates by analyzing the data from a longitudinal perspective. Data were collected at 3, 5, 9, and 12 months. Right censoring also occurred at these time points. Let $L_t$ denote covariates measured at time $t$, $A_0$ denote baseline treatment assignment, and $A_1$ through $A_4$ denote censoring nodes along the timeline (Figure 6). The set $\{A_0, A_1, A_2, A_3, A_4\}$ is referred to as the set of *intervention nodes*, because the causal contrast of interest involves intervening to assign a specific counterfactual treatment, and to ensure there is no right censoring.

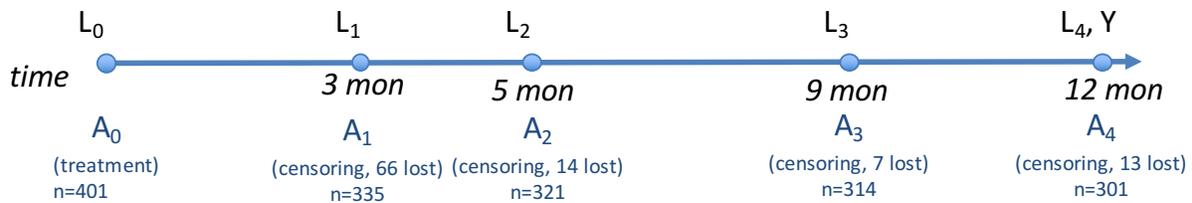

Figure 6. Schematic of data acquisition over time, beginning with treatment at time $t_0$, and right censoring occurring at time points t = 3, 5, 9, 12 months.

*5.2.1 Applying the TL Roadmap*

Step 1. *Statistical model:* All distributions of the data consistent with study inclusion/exclusion criteria, and respecting the process that gave rise to the data, e.g., treatment assignment was randomly assigned conditional on baseline covariates.

Step 2. *Causal model and causal parameter:* The causal parameter of interest remains the effect of randomization to acupuncture ($A_0 = 1$) vs no acupuncture ($A_0 = 0$), under no right censoring. The target parameter is defined as, $\psi_{ATE}^{causal} = E(Y_{11111}) - E(Y_{01111})$, in a causal model where baseline covariates, treatment, intercurrent events, time−varying covariates, and the outcome are expressly represented. The subscripts denote the counterfactual settings for each intervention node (baseline treatment, $A_0$, and missingness indicators). This definition of the parameter explicitly acknowledges the potential for intercurrent events at some time $t > 0$, to introduce bias. The causal model again assumes exogenous errors, and only those independencies implied by the time ordering.

Step 3. *Statistical Estimand and identifiability:* The statistical estimand is given by $\psi_{ATE}^{obs} = \bar{Q}^{\bar{a}_1} - \bar{Q}^{\bar{a}_1}$, where $\bar{a}_1$ is the longitudinal intervention that sets the intervention nodes to $(1,1,1,1,1)$, and $\bar{a}_0$ is the longitudinal intervention that sets the intervention nodes to $(0,1,1,1,1)$. By convention, $\bar{A}_t$ denotes the intervention node settings through time $t$, and $\bar{L}_t$ denotes the covariate history through time $t$. $\bar{Q}^{\bar{a}_1}$ and $\bar{Q}^{\bar{a}_0}$ can each be expressed as a series of nested conditional means, $\bar{Q}^{\bar{a}} = (\bar{Q}_Y^{\bar{a}}, \bar{Q}_{L_4}^{\bar{a}}, \bar{Q}_{L_3}^{\bar{a}}, \bar{Q}_{L_2}^{\bar{a}}, \bar{Q}_{L_1}^{\bar{a}}, \bar{Q}_{L_0}^{\bar{a}})$ (Bang and Robins 2005, Petersen et al. 2014),

$$\bar{Q}_Y^{\bar{a}} = E(Y_{\bar{a}}|\bar{L}_4, \bar{A}_4 = \bar{a}_4),$$
$$\bar{Q}_{L_4}^{\bar{a}} = E(\bar{Q}_Y^{\bar{a}}|\bar{L}_3, \bar{A}_3 = \bar{a}_3),$$
$$\bar{Q}_{L_3}^{\bar{a}} = E(\bar{Q}_{L_4}^{\bar{a}}|\bar{L}_2, \bar{A}_2 = \bar{a}_2),$$
$$\bar{Q}_{L_2}^{\bar{a}} = E(\bar{Q}_{L_3}^{\bar{a}}|\bar{L}_1, \bar{A}_1 = \bar{a}_1),$$
$$\bar{Q}_{L_1}^{\bar{a}} = E(\bar{Q}_{L_2}^{\bar{a}}|L_0, A_0 = a_0),$$
$$\bar{Q}^{\bar{a}} = E(\bar{Q}_{L_1}^{\bar{a}}).$$

The sequential positivity assumption must hold at all timepoints. Baseline treatment randomization guarantees positivity holds at $t = 0$, but not necessarily at subsequent timepoints. If right censoring differentially depletes a large proportion of observations within strata of confounders by trial arm, then a practical violation of the positivity assumption can occur. Time-varying PS diagnostics can provide helpful information for assessing whether this assumption is met. For the ITT analysis the consistency assumption is easily satisfied, since adherence to assigned treatment is immaterial. The CAR assumption is satisfied with respect to treatment assignment but is violated if there are unmeasured confounders of the associations between treatment, right censoring, and the outcome.

Step 4. *Estimation:* The statistical parameter can be estimated with longitudinal versions of IPTW (L-IPTW) and TMLE (L-TMLE), available in the *ltmle* R package (v. 1.2-0) (Lendle et al. 2017). Both estimators require modeling the PS and conditional probabilities of being observed at each intervention node, given the past ($G_t$). L-TMLE evaluates a nested conditional mean outcome regression at each time point, targeted towards the parameter of interest based on the cumulative product of the components of $G$ from $t_0$ through $t$. The longitudinal versions of IPTCW (L-IPTCW) at time point $t$ is the cumulative product of the components of $G$ from $t_0$ through $t$. Weights were truncated using the sample size-based strategy, $wt_{max} = \sqrt{n} \ln(n) / 5$. Plugging $n = 301$ into the formula (the number of observations that contribute to the estimation procedure at the last time point) yielded a maximum weight of 20.

*Modeling the missingness probabilities*: At each censoring node, the goal is to model the conditional probability of remaining uncensored through time $t$, given the past, $P(A_t = 1 \mid \bar{A}_{1:t-1} = 1, \bar{L}_t, A_0)$, where $\bar{A}_{1:t-1} = 1$ denotes remaining uncensored at all censoring intervention notes from 1 though $t$-$1$, $\bar{L}_t$ denotes baseline covariates and all time-varying covariates through time $t$, and $A_0$ is the initial treatment.

Downstream effects of treatment are potential causes of right censoring. Failing to adjust for time-dependent confounding would lead to residual bias in the effect estimate. However, for these data, adjusting for the many covariates and treatment recorded from baseline through time $t$ is complicated by the small number of censoring events at months 5, 9, and 12 ($n_{censored,t} = 14, 7, 13$, respectively). At these time points, regressing a rare binary outcome on a large set of covariates runs the risk of overfitting the model, even for methods that incorporate regularization (Friedman et al. 2001). For this reason, we designed the following two approaches:

> *Approach 1*. At 3 months 66 subjects were censored. We used SL to model $P(A_1 = 1 \mid A_0, \tilde{L}_0)$, where $\tilde{L}_0$ includes the same 15 baseline covariates considered in the data analyses presented in the prior analysis. For the remaining time points, where censoring events were extremely rare, we only adjusted for a single covariate, the initial treatment, by modeling $P(A_t = 1 \mid A_0)$, for $t > 1$.

> *Approach 2*. $P(A_1 = 1 \mid A_0, \tilde{L}_0)$ was modeled the same as in Approach 1. At each subsequent time point, we adjusted for treatment and time-dependent covariates by creating a one-dimensional summary score, $X_t$, as follows. SL was used to model the relationships between the outcome observed among the subjects who remained uncensored at $t = 4$, conditional on variables measured through time $t - 1$. These fitted models were used to predict $X_t$ for each subject who remained uncensored through $t - 1$. Logistic regression was then used to estimate $P(A_{t,t>1} = 1 \mid X_t)$.

The L-TMLE incorporated these SL-based estimates of $G_t$, and utilized the equivalent sample size-based truncation strategy (lower bound on each cumulative product term = 0.05). SL was also used to model the conditional expectations of the outcome given the past at each time point, working backwards from $t_4$ towards $t_0$.

Step 5. *Sensitivity analyses and substantive conclusion:* Potential violations of the identifying causal assumptions discussed in Step 3 cast doubt on the validity of a causal interpretation of the findings. A non-parametric sensitivity analysis will investigate their impact with respect to the magnitude of the presumed causal gap.

*5.2.2 Results*

Point estimates and 95% CI indicate that acupuncture reduces MWHS (Table 6). Estimates obtained using Approach 1 are quite similar to findings from the point-treatment analysis. Given that Approach 1 ignores time-varying covariates, this finding is not surprising. Incorporating time-varying data using Approach 2 shifted the L-IPTW point estimate further away from the null, with a nearly 10% increase in the SE. The L-TMLE point estimate were slightly shifted further away from the null, with little change to the SE and CI widths.

It may be that incorporating time-varying information had little impact on the point estimates because the majority of right censoring occurred at $t = 1$, before time-varying contributors to LTFU had manifested. In these analyses L-IPTW estimators were more variable than their point-treatment counterparts. TMLE estimates, L-TMLE estimates, and SEs were stable across all the analytic specifications, and had the narrowest CI.

Table 6. Results of longitudinal analyses using two approaches for modeling time-varying missingness probabilities, estimated ATE (Est), standard error (SE), 95% confidence interval lower bound (lb) and upper bound (ub), and p-value.

| Estimator | Approach | Est | SE | 95% CI lb | 95% CI ub | *p*-value |
|---|---|---|---|---|---|---|
| L-IPTW | 1 | -5.15 | 1.75 | -8.58 | -1.72 | 0.0040 |
|  | 2 | -5.64 | 1.93 | -9.43 | -1.85 | 0.0039 |
| L-TMLE | 1 | -5.23 | 1.17 | -7.53 | -2.93 | <0.0001 |
|  | 2 | -5.42 | 1.20 | -7.78 | -3.09 | <0.0001 |

This analysis highlights the value of capturing time-varying covariates related to treatment, the outcome, and the reason for withdrawing from the study. Instead of being forced to speculate on the magnitude of bias due to time-varying confounding, the longitudinal analysis provided the opportunity to evaluate its impact, at least with respect to measured potential confounders.

## 6. Discussion

TL provides a rigorous framework for learning from data, estimating causal effects, and evaluating their reliability. Following the roadmap ensures the statistical estimation problem is clearly defined and assumptions required for a causal interpretation are explicitly stated. TMLE+SL are rigorous statistical methodologies for addressing challenges inherent in analyses of studies that deviate from the ideal RCT, including those that incorporate RWD. Nonetheless, it is always preferable to minimize foreseeable challenges in the study design phase where possible, e.g., choosing a highly comparable external control arm, or devising strategies for mitigating treatment nonadherence. One should also plan at the outset to collect time-varying covariates, including reason for dropout. Some inherent challenges may not be avoidable, e.g., high-dimensional data, or rare outcome events. Non-parametric sensitivity analyses complement traditional parametric modeling approaches to evaluate the reliability of the RWE.

Our comparison of TMLE+SL with PS-based estimators shows the wisdom of defaulting to TMLE+SL in the planning stage. When challenges fail to arise, as in the setting of Case Study 1, most methods, including TMLE+SL, will be suitable. Introducing artificial selection bias and LTFU revealed that TMLE and IPTCW-SL were more successful than any PS matching variant under consideration, with TMLE out-performing IPTCW as the proportion LTFU increased. In Case Study 2, a similar pattern was observed in the point treatment analysis. In the longitudinal analysis, adjusting for baseline and time varying covariates made L-IPTW and L-TMLE point estimates more extreme than those based on baseline covariates alone. PS matching is not suitable for analyzing longitudinal data, so was not considered. The longitudinal analysis also illustrated a useful approach to reducing dimensionality when censoring events are rare, namely, creating a univariate summary of covariates measured through time $t$.

In the point-treatment setting, PS-Matching was often biased and more variable than IPW or TMLE. Post-match adjustment offered clear improvement, but an inability to handle LTFU or treatment non-adherence made it the weakest of the three approaches examined. IPTW, IPTCW (and L-IPTW) point estimates were in good agreement with TMLE (L-TMLE), but typically had larger SE and wider 95% CI. The loss in precision of the IPW estimator means that a larger sample size would be required for a study using IP weighting than one using TMLE to achieve the same power. Powering a study during the design phase should always be conservative, but the efficiency gain from using TMLE instead of IPW may be quantified in advance when similar external or pilot study data are available.

Trust in RWE should not be extended lightly, but we have shown how, with TL, trust can be earned when warranted. Taken together, the work products, the definition of the statistical model, causal parameter, statistical estimand, discussion of plausibility of identifying assumptions, point estimates, SE, p-values, CI, PS diagnostics, and sensitivity analyses, paint a full picture. They provide a rational basis for helping to decide whether to either accept the substantive conclusion drawn from the study findings, or to postpone acceptance until alternative information upon which to base a firm conclusion can be obtained.

## References

Balzer L.B., Zheng W., van der Laan M.J., and Petersen M.L. (2019). A new approach to hierarchical data analysis: Targeted maximum likelihood estimation for the causal effect of a cluster-level exposure. *Statistical Methods in Medical Research*, 28(6), 1761–1780. https://doi.org/10.1177/0962280218774936.

Bang, H. and Robins, J.M. (2005), Doubly Robust Estimation in Missing Data and Causal Inference Models. *Biometrics*, 61, 962-973. https://doi.org/10.1111/j.1541-0420.2005.00377.x.
Carrell D., Gruber S., Floyd J.S., Bann M.A., Cushing-Haugen K.L., Johnson R.L., Graham V., Cronkite D.J., Hazlehurst B.L., Felcher A.H., Bejin C.A., Kennedy A., Shinde M., Karami S., Ma Y., Stojanovic D., Zhao Y., Ball R., Nelson J. (2022). Improving Methods of Identifying Anaphylaxis for Medical Product Safety Surveillance Using Natural Language Processing and Machine Learning. *American Journal of Epidemiology*, in press.

Chipman H., George E.I., and McCulloch R.E. (2010). BART: Bayesian additive regression trees. *The Annals of Applied Statistics*, 4(1), 266-298. doi: 10.1214/09-AOAS285.

Cole S.R., Hernán M.A., Constructing Inverse Probability Weights for Marginal Structural Models. *American Journal of Epidemiology*, 168(6), 656–664, https://doi.org/10.1093/aje/kwn164.

Corrigan-Curay, J., Sacks, L., and Woodcock, J (2018). Real-world evidence and real-world data for evaluating drug safety and effectiveness. *Journal of the American Medical Association*, 320, 867–868.

Díaz I., & van der Laan M. J. (2013). Sensitivity analysis for causal inference under unmeasured confounding and measurement error problems. *The International Journal of Biostatistics*, 9(2), 149-160.

FDA (2018), "Framework for FDA's Real-world Evidence Program," Available at https://www.fda.gov/media/120060/download.
FDA (2020) " 21st Century Cures Act," Available at https://www.fda.gov/regulatory-information/selected-amendments-fdc-act/21st-century-cures-act

FDA (2021), "Real-World Data: Assessing Electronic Health Records and Medical Claims Data To Support Regulatory Decision-Making for Drug and Biological Products," Available athttps://www.fda.gov/regulatory-information/search-fda-guidance-documents/real-world-data-assessing-electronic-health-records-and-medical-claims-data-support-regulatory.

Franklin, J.M., Schneeweiss, S., Polinski, J. M., & Rassen, J.A. (2014). Plasmode simulation for the evaluation of pharmacoepidemiologic methods in complex healthcare databases. *Computational Statistics & Data Analysis*, 72, 219–226. https://doi.org/10.1016/j.csda.2013.10.018

Friedman J., Hastie T., and Tibshirani, R. (2001). *The Elements of Statistical Learning*. New York: Springer series in statistics.

Friedman J., Hastie T., Tibshirani R. (2010). Regularization Paths for Generalized Linear Models via Coordinate Descent. *Journal of Statistical Software*, 33(1), 1-22.

Gruber, S., and van der Laan, M. (2012). tmle: An R Package for Targeted Maximum Likelihood Estimation. *Journal of Statistical Software*, 51(13), 1–35. https://doi.org/10.18637/jss.v051.i13.

Gruber S., Krakower D., Menchaca J.T., Hsu K., Hawrusik R., Maro J.C., Cocoros N.M., Kruskal B.A., Wilson I.B., Mayer K.H., Klompas M (2020). Using electronic health records to identify candidates for human immunodeficiency virus pre-exposure prophylaxis: An application of super learning to risk prediction when the outcome is rare. *Statistics in Medicine*, 39(23),3059-3073. doi: 10.1002/sim.8591.

Gruber S., Phillips R., Lee H., and van der Laan M. (2022). Data-adaptive selection of the propensity score truncation level for inverse probability weighted and targeted maximum likelihood estimators of marginal point treatment effects. *American Journal of Epidemiology*, kwac087, https://doi.org/10.1093/aje/kwac087.

Hansen B.B. (2004). Full Matching in an Observational Study of Coaching for the SAT. *Journal of the American Statistical Association*, 99(467), 609-618, DOI: 10.1198/016214504000000647.

Hastie T. (2019). gam: Generalized Additive Models. R package version 1.16.1. https://CRAN.R-project.org/package=gam.

Hernán M.A., Brumback B., and Robins J.M. (2000). Marginal structural models to estimate the causal effect of zidovudine on the survival of HIV-positive men. *Epidemiology*, 11(5), 561–570.

Ho D.E., Imai K., King G., Stuart E.A. (2011). MatchIt: Nonparametric Preprocessing for Parametric Causal Inference. *Journal of Statistical Software*, 42(8), 1–28. DOI: 10.18637/jss.
Ho M., van der Laan M., Lee H., Chen J., Lee K., Fang Y., He W., Irony T., Jiang Q., Lin X., Meng Z., Mishra-Kalyani P., Rockhold F., Song Y., Wang H., and White R. (2021) The Current Landscape in Biostatistics of Real-World Data and Evidence: Causal Inference Frameworks for Study Design and Analysis. *Statistics in Biopharmaceutical Research*, 1-14. DOI:10.1080/19466315.2021.1883475

ICH (2020), "ICH E9(R1) Addendum to Statistical Principles for Clinical Trials on Choosing Appropriate Estimands and Defining Sensitivity Analyses in Clinical Trials," Available at https://www.ich.org/page/efficacy-guidelines.

Imai K. and Ratkovic M. (2013). Covariate balancing propensity score. Covariate balancing propensity score. *Journal of the Royal Statistical Society Series B*, 76, 243-263, https://doi.org/10.1111/rssb.12027

International Stroke Trial Collaborative Group (1997). The international stroke trial (IST): a randomised trial of aspirin, subcutaneous heparin, both, or neither among 19435 patients with acute ischaemic stroke. *The Lancet*, 349,1569–1581.

Kahale L.A., Diab B., Khamis A.M., Chang Y., Cruz Lopes L., Agarwal A., Li L., Mustafa R.A., Koujanian S., Waziry R., Busse J.W., Dakik A., Guyatt G., and Aki E.A. (2019). Potentially missing data are considerably more frequent than definitely missing data: a methodological survey of 638 randomized controlled trials. *Journal of Clinical Epidemiology*, 106, 18–31.

Kempker R. R., Mikiashvili L., Zhao Y., Benkeser D., Barbakadze K., Bablishvili N., Avaliani Z., Peloquin C.A., Blumberg H.M., and Kipiani M. (2020). Clinical Outcomes Among Patients With Drug-resistant Tuberculosis Receiving Bedaquiline- or Delamanid-Containing Regimens. *Clinical Infectious Diseases : an official publication of the Infectious Diseases Society of America*, 71(9), 2336–2344. https://doi.org/10.1093/cid/ciz1107.

Kreif N., Tran L., Grieve R., De Stavola B., Tasker R. C., and Petersen M. (2017). Estimating the Comparative Effectiveness of Feeding Interventions in the Pediatric Intensive Care Unit: A Demonstration of Longitudinal Targeted Maximum Likelihood Estimation. *American Journal of Epidemiology*, 186(12), 1370–1379. https://doi.org/10.1093/aje/kwx213.

Lendle, S. D., Fireman, B., and van der Laan, M. J. (2013). Targeted maximum likelihood estimation in safety analysis. *Journal of Clinical Epidemiology*, 66(8 Suppl), S91–S98. https://doi.org/10.1016/j.jclinepi.2013.02.017.

Lendle S.D., Schwab J., Petersen M.L., and van der Laan M.J. (2017). ltmle: An R Package Implementing Targeted Minimum Loss-Based Estimation for Longitudinal Data. *Journal of Statistical Software*, 81(1), 1-21. doi:10.18637/jss.v081.i01 (URL: https://doi.org/10.18637/jss.v081.i01).

Ling A.Y., Montez-Rath M. E., Mathur M. B., Kapphahn K., and Desai M. (2019). How to apply multiple imputation in propensity score matching with partially observed confounders: a simulation study and practical recommendations. arXiv preprint arXiv:1904.07408.

Petersen M.L., and van der Laan M.J. (2014). Causal Models and Learning from Data: Integrating Causal Modeling and Statistical Estimation. *Epidemiology* (Cambridge, Mass.), 25, 418. https://doi.org/10.1097/EDE.0000000000000078.

Petersen M., Schwab J., Gruber S., Blaser N., Schomaker M., and van der Laan M (2014). Targeted maximum likelihood estimation for dynamic and static longitudinal marginal structural working models. *Journal of Causal Inference*, 2(2), 147-185.

Pirrachio R., Petersen M.L., Carone M., Reche Rigon M., Chrette S., and van der Laan M.J. (2015). Mortality prediction in intensive care units with the Super ICU Learner Algorithm (SICULA): a population-based study. *The Lancet Respiratory Medicine*, 3(1), 42–52.

Polley E., LeDell E., Kennedy C., and van der Laan, M. (2019). SuperLearner: Super Learner Prediction. R package version 2.0-26. https://CRAN.R-project.org/package=SuperLearner.

Porter K.E., Gruber S., van der Laan M.J., and Sekhon J.S. (2011). The Relative Performance of Targeted Maximum Likelihood Estimators. *The International Journal of Biostatistics*, 7(1), Article 31.

R Core Team (2020). R: A language and environment for statistical computing. R Foundation for Statistical Computing, Vienna, Austria. URL http://www.R-project.org/.

Rosenbaum P.R. and Rubin D.B. (1983). The central role of the propensity score in observational studies for causal effects. *Biometrika*, 70, 41–55.

Sandercock P., Niewada M., and Czlonkowska A.(2011). International Stroke Trial database (version 2), University of Edinburgh. Department of Clinical Neurosciences. doi.org/10.7488/ds/104.

Stuart E.A. (2010). Matching methods for causal inference: A review and a look forward. *Statistical science : A review journal of the Institute of Mathematical Statistics*, 25(1), 1–21. https://doi.org/10.1214/09-STS313.

van der Laan M. J., and Rose S. (2011), *Targeted Learning: Causal Inference for Observational and Experimental Data*, New York: Springer.

Vickers A.J., McCarney R. (2003). Use of a single global assessment to reduce missing data in a clinical trial with follow-up at one year. *Controlled clinical trials*, 24(6), 731-735. https://doi.org/10.10.16/j.cct.2003.10.001.

Vickers A.J., Rees R.W., Zollman C.E., McCarney R., Smith C., Ellis N., Fisher P., and Van Haselen R. (2004). Acupuncture for chronic headache in primary care: large, pragmatic, randomised trial. *BMJ*. doi: 10.1136/bmj.38029.421863.EB.

Vickers A.J. (2006). Whose data set is it anyway? sharing raw data from randomized trials. *Trials*, 7. doi: doi:10.1186/1745-6215-7-15.

Zeileis A. (2004). Econometric Computing with HC and HAC Covariance Matrix Estimators. *Journal of Statistical Software*, 11(10), 1-17. doi:10.18637/jss.v011.i10.

Zeileis A., Köll S., Graham N. (2020). Various Versatile Variances: An Object-Oriented Implementation of Clustered Covariances in R. *Journal of Statistical Software*, 95(1), 1-36. doi: 10.18637/jss.v095.i01.

# Appendix

Table A1. N=24 covariates available in the IST dataset; 466 hospitals in 36 countries took part in the RCT. Hospitals and countries were grouped by geographical region, based on World Bank groupings: East Asia; Pacific Europe and Central Asia; Latin America and Caribbean; Middle East and North Africa; North America; South Asia; Sub-Saharan Africa (World Bank 2021).

Table A1. Task 1 baseline covariates included in the International Stroke Trial dataset.

| Variable Name | Description |
| --- | --- |
| RDELAY | Delay between stroke and randomization in hours |
| RCONSC | Conscious state at randomization (F - fully alert, D - drowsy, U - unconscious) |
| SEX | M=male, F=female |
| AGE | Age in years |
| RSLEEP | Symptoms noted on waking (Y/N) |
| RATRIAL[a] | Atrial fibrillation (Y/N) |
| RCT | CT before randomization (Y/N) |
| RVISINF | Infarct visible on CT (Y/N) |
| RHEP24[a] | Heparin within 24 hours prior to randomization (Y/N) |
| RASP3[a] | Aspirin within 3 days prior to randomization (Y/N) |
| RSBP | Systolic blood pressure at randomization (mmHg) |
| RDEF1 | Face deficit (Y/N/C=can't assess) |
| RDEF2 | Arm/hand deficit (Y/N/C=can't assess) |
| RDEF3 | Leg/foot deficit (Y/N/C=can't assess) |
| RDEF4 | Dysphasia (Y/N/C=can't assess) |
| RDEF5 | Hemianopia (Y/N/C=can't assess) |
| RDEF6 | Visuospatial disorder (Y/N/C=can't assess) |
| RDEF7 | Brainstem/cerebellar signs (Y/N/C=can't assess) |
| RDEF8 | Other deficit (Y/N/C=can't assess) |
| STYPE | Stroke subtype (TACS/PACS/POCS/LACS/other) |
| RXHEP | Trial heparin allocated (M/L/N) |
| MISSING1[a] | Indicator of imputed value of "U" for RATRIAL and RASP3 |
| MISSING2[a] | Indicator of imputed value of "U" for RHEP24 |
| REGION | Geographical region |

[a] Missing values were recoded to "U" to indicate unknown for RATRIAL and RASP3 (984 pilot phase subjects), or RHEP24 (344 values).

Table A2. IST trial: standardized mean differences for all covariates, before and after matching and IP weighting

| Covariate | Before IP Weighting or Matching | | | | | | After Full Matching | | | | | | After IP Weighting | | | | | |
|---|---|---|---|---|---|---|---|---|---|---|---|---|---|---|---|---|---|---|
| | Controls (n = 9,699) | | Treated (n = 9,703) | | SMD | p-val | Controls (n = 9,699) | | Treated (n = 9,703) | | SMD | p-val | Controls (n = 9,699) | | Treated (n = 9,703) | | SMD | p-val |
| | mean | SD | mean | SD | | | mean | SD | mean | SD | | | mean | SD | mean | SD | | |
| Age | 71.7 | 11.6 | 71.72 | 11.64 | 0 | 1 | 71.64 | 11.7 | 71.72 | 11.64 | 0.01 | 1 | 71.71 | 11.62 | 71.71 | 11.62 | 0 | 1 |
| Number hours from initial stroke to randomization | 20.11 | 12.52 | 20.13 | 12.42 | 0 | 1 | 20.16 | 12.51 | 20.13 | 12.42 | 0 | 1 | 20.12 | 12.53 | 20.12 | 12.42 | 0 | 1 |
| Systolic blood pressure | 160.32 | 27.52 | 160.02 | 27.7 | 0.01 | 1 | 160.01 | 27.4 | 160.02 | 27.7 | 0 | 1 | 160.18 | 27.5 | 160.18 | 27.74 | 0 | 1 |
| | counts | prop | counts | prop | | | counts | prop | counts | prop | | | counts | prop | counts | prop | | |
| Missing atrial fibrillation, aspirin within 3 days prior to randomization | 491 | 5.10% | 493 | 5.10% | 0 | 1 | 493.42 | 5.10% | 493 | 5.10% | 0 | 1 | 491.96 | 5.10% | 492.08 | 5.10% | 0 | 1 |
| Missing heparin within 24 hours prior to randomization | 172 | 1.80% | 172 | 1.80% | 0 | 1 | 179.41 | 1.80% | 172 | 1.80% | 0 | 1 | 172.12 | 1.80% | 172.18 | 1.80% | 0 | 1 |
| Aspirin within 3 days prior to randomization (N) | 7233 | 74.60% | 7252 | 74.70% | 0 | 1 | 7241.42 | 74.70% | 7252 | 74.70% | 0 | 1 | 7241.59 | 74.70% | 7244.58 | 74.70% | 0 | 1 |
| Aspirin within 3 days prior to randomization (unknown) | 491 | 5.10% | 493 | 5.10% | 0 | 1 | 493.42 | 5.10% | 493 | 5.10% | 0 | 1 | 491.96 | 5.10% | 492.08 | 5.10% | 0 | 1 |
| Aspirin within 3 days prior to randomization (Y) | 1975 | 20.40% | 1958 | 20.20% | 0 | 1 | 1964.16 | 20.30% | 1958 | 20.20% | 0 | 1 | 1965.4 | 20.30% | 1966.39 | 20.30% | 0 | 1 |
| Atrial fibrillation (N) | 7662 | 79% | 7594 | 78.30% | 0.01 | 1 | 7614.54 | 78.50% | 7594 | 78.30% | 0 | 1 | 7625.88 | 78.60% | 7629.18 | 78.60% | 0 | 1 |
| Atrial fibrillation (unknown) | 491 | 5.10% | 493 | 5.10% | 0 | 1 | 493.42 | 5.10% | 493 | 5.10% | 0 | 1 | 491.96 | 5.10% | 492.08 | 5.10% | 0 | 1 |
| Atrial fibrillation (Y) | 1546 | 15.90% | 1616 | 16.70% | 0.01 | 1 | 1591.04 | 16.40% | 1616 | 16.70% | 0 | 1 | 1581.12 | 16.30% | 1581.79 | 16.30% | 0 | 1 |
| Conscious state at randomization (drowsy) | 2121 | 21.90% | 2123 | 21.90% | 0 | 1 | 2121.28 | 21.90% | 2123 | 21.90% | 0 | 1 | 2120.8 | 21.90% | 2121.68 | 21.90% | 0 | 1 |
| Conscious state at randomization (fully alert) | 7448 | 76.80% | 7451 | 76.80% | 0 | 1 | 7438.19 | 76.70% | 7451 | 76.80% | 0 | 1 | 7448.77 | 76.80% | 7451.9 | 76.80% | 0 | 1 |
| Conscious state at randomization (unconscious) | 130 | 1.30% | 129 | 1.30% | 0 | 1 | 139.53 | 1.40% | 129 | 1.30% | 0 | 1 | 129.38 | 1.30% | 129.46 | 1.30% | 0 | 1 |
| CT before randomization (N) | 3175 | 32.70% | 3222 | 33.20% | 0.01 | 1 | 3207.95 | 33.10% | 3222 | 33.20% | 0 | 1 | 3198.69 | 33% | 3199.9 | 33% | 0 | 1 |
| CT before randomization (Y) | 6524 | 67.30% | 6481 | 66.80% | 0.01 | 1 | 6491.05 | 66.90% | 6481 | 66.80% | 0 | 1 | 6500.26 | 67% | 6503.15 | 67% | 0 | 1 |
| Face deficit (can't asses) | 124 | 1.30% | 122 | 1.30% | 0 | 1 | 114.43 | 1.20% | 122 | 1.30% | 0 | 1 | 122.32 | 1.30% | 122.3 | 1.30% | 0 | 1 |

| | | | | | | | | | | | | | | | | |
|---|---|---|---|---|---|---|---|---|---|---|---|---|---|---|---|---|
| Face deficit (N) | 2515 | 25.90% | 2563 | 26.40% | 0.01 | 1 | 2513.51 | 25.90% | 2563 | 26.40% | 0.01 | 1 | 2537.89 | 26.20% | 2538.95 | 26.20% | 0 | 1 |
| Face deficit (Y) | 7060 | 72.80% | 7018 | 72.30% | 0.01 | 1 | 7071.06 | 72.90% | 7018 | 72.30% | 0.01 | 1 | 7038.74 | 72.60% | 7041.8 | 72.60% | 0 | 1 |
| Arm/hand deficit (can't assess) | 68 | 0.70% | 55 | 0.60% | 0 | 1 | 53.4 | 0.60% | 55 | 0.60% | 0 | 1 | 61.68 | 0.60% | 61.77 | 0.60% | 0 | 1 |
| Arm/hand deficit (N) | 1317 | 13.60% | 1345 | 13.90% | 0 | 1 | 1307.29 | 13.50% | 1345 | 13.90% | 0.01 | 1 | 1331.22 | 13.70% | 1331.7 | 13.70% | 0 | 1 |
| Arm/hand deficit (Y) | 8314 | 85.70% | 8303 | 85.60% | 0 | 1 | 8338.31 | 86% | 8303 | 85.60% | 0 | 1 | 8306.05 | 85.60% | 8309.57 | 85.60% | 0 | 1 |
| Leg/foot deficit (can't assess) | 122 | 1.30% | 133 | 1.40% | 0 | 1 | 124.41 | 1.30% | 133 | 1.40% | 0 | 1 | 127.46 | 1.30% | 127.64 | 1.30% | 0 | 1 |
| Leg/foot deficit (N) | 2225 | 22.90% | 2268 | 23.40% | 0.01 | 1 | 2234.4 | 23% | 2268 | 23.40% | 0 | 1 | 2246.05 | 23.20% | 2247.05 | 23.20% | 0 | 1 |
| Leg/foot deficit (Y) | 7352 | 75.80% | 7302 | 75.30% | 0.01 | 1 | 7340.19 | 75.70% | 7302 | 75.30% | 0 | 1 | 7325.43 | 75.50% | 7328.36 | 75.50% | 0 | 1 |
| Dysphasia (can't assess) | 303 | 3.10% | 280 | 2.90% | 0.01 | 1 | 285.53 | 2.90% | 280 | 2.90% | 0 | 1 | 291.31 | 3% | 291.17 | 3% | 0 | 1 |
| Dysphasia (N) | 5160 | 53.20% | 5164 | 53.20% | 0 | 1 | 5191.65 | 53.50% | 5164 | 53.20% | 0 | 1 | 5162.65 | 53.20% | 5164.74 | 53.20% | 0 | 1 |
| Dysphasia (Y) | 4236 | 43.70% | 4259 | 43.90% | 0 | 1 | 4221.82 | 43.50% | 4259 | 43.90% | 0 | 1 | 4244.99 | 43.80% | 4247.13 | 43.80% | 0 | 1 |
| Hemianopia (can't assess) | 1955 | 20.20% | 1980 | 20.40% | 0 | 1 | 1966.59 | 20.30% | 1980 | 20.40% | 0 | 1 | 1966.31 | 20.30% | 1967.11 | 20.30% | 0 | 1 |
| Hemianopia (N) | 6186 | 63.80% | 6187 | 63.80% | 0 | 1 | 6153.98 | 63.40% | 6187 | 63.80% | 0 | 1 | 6184.92 | 63.80% | 6187.59 | 63.80% | 0 | 1 |
| Hemianopia (Y) | 1558 | 16.10% | 1536 | 15.80% | 0 | 1 | 1578.43 | 16.30% | 1536 | 15.80% | 0.01 | 1 | 1547.72 | 16% | 1548.34 | 16% | 0 | 1 |
| Visuospatial disorder (can't assess) | 1733 | 17.90% | 1712 | 17.60% | 0 | 1 | 1717.07 | 17.70% | 1712 | 17.60% | 0 | 1 | 1722.66 | 17.80% | 1723.39 | 17.80% | 0 | 1 |
| Visuospatial disorder (N) | 6382 | 65.80% | 6406 | 66% | 0 | 1 | 6368.2 | 65.70% | 6406 | 66% | 0 | 1 | 6392.31 | 65.90% | 6394.91 | 65.90% | 0 | 1 |
| Visuospatial disorder (Y) | 1584 | 16.30% | 1585 | 16.30% | 0 | 1 | 1613.73 | 16.60% | 1585 | 16.30% | 0 | 1 | 1583.98 | 16.30% | 1584.74 | 16.30% | 0 | 1 |
| Brainstem/cerebellar signs (can't assess) | 795 | 8.20% | 794 | 8.20% | 0 | 1 | 795.05 | 8.20% | 794 | 8.20% | 0 | 1 | 794.25 | 8.20% | 794.54 | 8.20% | 0 | 1 |
| Brainstem/cerebellar signs (N) | 7843 | 80.90% | 7835 | 80.70% | 0 | 1 | 7831.85 | 80.70% | 7835 | 80.70% | 0 | 1 | 7837.76 | 80.80% | 7841.19 | 80.80% | 0 | 1 |
| Brainstem/cerebellar signs (Y) | 1061 | 10.90% | 1074 | 11.10% | 0 | 1 | 1072.1 | 11.10% | 1074 | 11.10% | 0 | 1 | 1066.94 | 11% | 1067.32 | 11% | 0 | 1 |
| Other deficit (can't assess) | 616 | 6.40% | 632 | 6.50% | 0 | 1 | 615.2 | 6.30% | 632 | 6.50% | 0 | 1 | 623.52 | 6.40% | 623.71 | 6.40% | 0 | 1 |
| Other deficit (N) | 8470 | 87.30% | 8466 | 87.30% | 0 | 1 | 8461.03 | 87.20% | 8466 | 87.30% | 0 | 1 | 8466.97 | 87.30% | 8470.66 | 87.30% | 0 | 1 |
| Other deficit (Y) | 613 | 6.30% | 605 | 6.20% | 0 | 1 | 622.77 | 6.40% | 605 | 6.20% | 0 | 1 | 608.46 | 6.30% | 608.68 | 6.30% | 0 | 1 |
| Region (East Asia and Pacific) | 657 | 6.80% | 656 | 6.80% | 0 | 1 | 657 | 6.80% | 656 | 6.80% | 0 | 1 | 655.69 | 6.80% | 655.99 | 6.80% | 0 | 1 |
| Region (Europe and Central Asia) | 8215 | 84.70% | 8217 | 84.70% | 0 | 1 | 8215 | 84.70% | 8217 | 84.70% | 0 | 1 | 8214.13 | 84.70% | 8217.52 | 84.70% | 0 | 1 |
| Region (Latin America and Caribbean) | 345 | 3.60% | 347 | 3.60% | 0 | 1 | 345 | 3.60% | 347 | 3.60% | 0 | 1 | 346.37 | 3.60% | 346.49 | 3.60% | 0 | 1 |
| Region (Middle East and North Africa) | 200 | 2.10% | 200 | 2.10% | 0 | 1 | 200 | 2.10% | 200 | 2.10% | 0 | 1 | 200.34 | 2.10% | 200.44 | 2.10% | 0 | 1 |
| Region (North America) | 124 | 1.30% | 124 | 1.30% | 0 | 1 | 124 | 1.30% | 124 | 1.30% | 0 | 1 | 123.79 | 1.30% | 123.91 | 1.30% | 0 | 1 |

| | | | | | | | | | | | | | | | | |
|---|---|---|---|---|---|---|---|---|---|---|---|---|---|---|---|---|
| Region (South Asia) | 124 | 1.30% | 124 | 1.30% | 0 | 1 | 124 | 1.30% | 124 | 1.30% | 0 | 1 | 124.01 | 1.30% | 124.07 | 1.30% | 0 | 1 |
| Region (Sub-Saharan Africa) | 34 | 0.40% | 35 | 0.40% | 0 | 1 | 34 | 0.40% | 35 | 0.40% | 0 | 1 | 34.62 | 0.40% | 34.63 | 0.40% | 0 | 1 |
| Heparin within 24 hours prior to randomization (N) | 9306 | 95.90% | 9316 | 96% | 0 | 1 | 9319.21 | 96.10% | 9316 | 96% | 0 | 1 | 9309.08 | 96% | 9313 | 96% | 0 | 1 |
| Heparin within 24 hours prior to randomization (unknown) | 172 | 1.80% | 172 | 1.80% | 0 | 1 | 179.41 | 1.80% | 172 | 1.80% | 0 | 1 | 172.12 | 1.80% | 172.18 | 1.80% | 0 | 1 |
| Heparin within 24 hours prior to randomization (Y) | 221 | 2.30% | 215 | 2.20% | 0 | 1 | 200.38 | 2.10% | 215 | 2.20% | 0 | 1 | 217.75 | 2.20% | 217.87 | 2.20% | 0 | 1 |
| Symptoms noted on waking (N) | 6839 | 70.50% | 6888 | 71% | 0.01 | 1 | 6854.33 | 70.70% | 6888 | 71% | 0 | 1 | 6863.26 | 70.80% | 6866.18 | 70.80% | 0 | 1 |
| Symptoms noted on waking (Y) | 2860 | 29.50% | 2815 | 29% | 0.01 | 1 | 2844.67 | 29.30% | 2815 | 29% | 0 | 1 | 2835.69 | 29.20% | 2836.87 | 29.20% | 0 | 1 |
| Infarct visible on CT (N) | 6468 | 66.70% | 6531 | 67.30% | 0.01 | 1 | 6490.8 | 66.90% | 6531 | 67.30% | 0 | 1 | 6499.58 | 67% | 6502.25 | 67% | 0 | 1 |
| Infarct visible on CT (Y) | 3231 | 33.30% | 3172 | 32.70% | 0.01 | 1 | 3208.2 | 33.10% | 3172 | 32.70% | 0.01 | 1 | 3199.37 | 33% | 3200.8 | 33% | 0 | 1 |
| Trial heparin allocated (high dose, pilot phase) | 122 | 1.30% | 123 | 1.30% | 0 | 1 | 119.78 | 1.20% | 123 | 1.30% | 0 | 1 | 122.58 | 1.30% | 122.6 | 1.30% | 0 | 1 |
| Trial heparin allocated (low dose) | 2427 | 25% | 2429 | 25% | 0 | 1 | 2418.44 | 24.90% | 2429 | 25% | 0 | 1 | 2426.76 | 25% | 2427.92 | 25% | 0 | 1 |
| Trial heparin allocated (high dose) | 2299 | 23.70% | 2301 | 23.70% | 0 | 1 | 2282 | 23.50% | 2301 | 23.70% | 0 | 1 | 2300.9 | 23.70% | 2301.8 | 23.70% | 0 | 1 |
| Trial heparin allocated (N) | 4851 | 50% | 4850 | 50% | 0 | 1 | 4878.77 | 50.30% | 4850 | 50% | 0 | 1 | 4848.71 | 50% | 4850.73 | 50% | 0 | 1 |
| Sex (F) | 4456 | 45.90% | 4559 | 47% | 0.01 | 1 | 4489.54 | 46.30% | 4559 | 47% | 0.01 | 1 | 4506.45 | 46.50% | 4508.4 | 46.50% | 0 | 1 |
| Sex (M) | 5243 | 54.10% | 5144 | 53% | 0.01 | 1 | 5209.46 | 53.70% | 5144 | 53% | 0.01 | 1 | 5192.5 | 53.50% | 5194.65 | 53.50% | 0 | 1 |
| Stroke subtype (lacunar) | 2325 | 24% | 2323 | 23.90% | 0 | 1 | 2294.79 | 23.70% | 2323 | 23.90% | 0 | 1 | 2323.67 | 24% | 2324.59 | 24% | 0 | 1 |
| Stroke subtype (other) | 31 | 0.30% | 26 | 0.30% | 0 | 1 | 26.37 | 0.30% | 26 | 0.30% | 0 | 1 | 28.51 | 0.30% | 28.52 | 0.30% | 0 | 1 |
| Stroke subtype (partial anterior circulation) | 3932 | 40.50% | 3912 | 40.30% | 0 | 1 | 3931.34 | 40.50% | 3912 | 40.30% | 0 | 1 | 3921.87 | 40.40% | 3923.61 | 40.40% | 0 | 1 |
| Stroke subtype (posterior circulation) | 1104 | 11.40% | 1120 | 11.50% | 0 | 1 | 1118.28 | 11.50% | 1120 | 11.50% | 0 | 1 | 1111.43 | 11.50% | 1111.83 | 11.50% | 0 | 1 |
| Stroke subtype (total anterior circulation) | 2307 | 23.80% | 2322 | 23.90% | 0 | 1 | 2328.22 | 24% | 2322 | 23.90% | 0 | 1 | 2313.47 | 23.90% | 2314.5 | 23.90% | 0 | 1 |

Table A3. Coefficients for covariates used to model the outcome for IST simulation study.

| Covariate | Coefficient | Covariate | Coefficient | Covariate | Coefficient |
|---|---|---|---|---|---|
| **RXHEP_L** | -0.0655 | **RDEF1_Y** | 0.1142 | **RDEF7_N** | -0.1835 |
| **RCONSC_D** | 0.9721 | **RDEF2_Y** | 0.2436 | **RDEF7_Y** | 0.2602 |
| **RCONSC_U** | 2.3167 | **RDEF3_N** | -0.2544 | **RDEF8_C** | -0.0870 |
| **RSLEEP_N** | 0.0523 | **RDEF3_Y** | 0.2703 | **RDEF8_Y** | 0.1889 |
| **RVISINF_N** | -0.2448 | **RDEF4_C** | 0.3684 | **STYPE_LACS** | -0.5917 |
| **RCT_N** | 0.2096 | **RDEF4_N** | -0.0810 | **STYPE_TACS** | 0.1611 |
| **RATRIAL_N** | -0.3407 | **RDEF5_C** | 0.0588 | **Z1** | 0.0330 |
| **RASP3_Y** | -0.1531 | **RDEF5_N** | -0.3180 | **Z2** | 0.8578 |
| **RHEP24_U** | -0.1604 | **RDEF6_C** | 0.1604 | **Z3** | -0.1628 |
| **RDEF1_N** | -0.1837 | **RDEF6_N** | -0.1591 | **Z4** | 1.2844 |

Table A4. IST Plasmode simulation study: Standardized mean differences in the observed data indicate that covariate imbalances are greatly reduced after matching or inverse weighting.

| Covariate | Before IP Weighting or Matching | | | | | | After Full Matching | | | | | | After IP Weighting | | | | | |
|---|---|---|---|---|---|---|---|---|---|---|---|---|---|---|---|---|---|---|
| | Controls (n = 16,042) | | Treated (n = 3,360) | | SMD | p-val | Controls (n = 16,042) | | Treated (n = 3,360) | | SMD | p-val | Controls (n = 16,042) | | Treated (n = 3,360) | | SMD | p-val |
| | mean | SD | mean | SD | | | mean | SD | mean | SD | | | mean | SD | mean | SD | | |
| Age | 71.67 | 11.59 | 71.91 | 11.74 | 0.02 | 0.98 | 71.5 | 11.97 | 71.91 | 11.74 | 0.03 | 0.97 | 71.7 | 11.63 | 71.62 | 11.68 | 0.01 | 0.99 |
| Number hours from initial stroke to randomization | 21.58 | 12.30 | 13.15 | 10.80 | 0.73 | 0.47 | 13.63 | 9.89 | 13.15 | 10.80 | 0.05 | 0.96 | 20.19 | 12.25 | 22.45 | 14.26 | 0.17 | 0.86 |
| Systolic blood pressure | 159.58 | 27.45 | 162.99 | 28.18 | 0.12 | 0.9 | 164.03 | 29.77 | 162.99 | 28.18 | 0.04 | 0.97 | 160.28 | 27.81 | 161.44 | 27.12 | 0.04 | 0.97 |
| | prop | SD | prop | SD | | | prop | SD | prop | SD | | | prop | SD | prop | SD | | |
| Missing atrial fibrillation, aspirin within 3 days prior to randomization | 5.11% | 0.48 | 4.88% | 0.47 | 0 | 1 | 2.26% | 0.39 | 4.88% | 0.47 | 0.07 | 0.95 | 5.06% | 0.47 | 4.25% | 0.45 | 0.02 | 0.99 |
| Missing heparin within 24 hours prior to randomization | 1.84% | 0.37 | 1.46% | 0.35 | 0.01 | 0.99 | 0.45% | 0.26 | 1.46% | 0.35 | 0.04 | 0.97 | 1.76% | 0.36 | 1.37% | 0.34 | 0.01 | 0.99 |
| Aspirin within 3 days prior to randomization (N) | 74.38% | 0.93 | 75.98% | 0.93 | 0.02 | 0.99 | 78.45% | 0.94 | 75.98% | 0.93 | 0.03 | 0.98 | 74.72% | 0.93 | 75.52% | 0.93 | 0.01 | 0.99 |
| Aspirin within 3 days prior to randomization (unknown) | 5.11% | 0.48 | 4.88% | 0.47 | 0 | 1 | 2.26% | 0.39 | 4.88% | 0.47 | 0.07 | 0.95 | 5.06% | 0.47 | 4.25% | 0.45 | 0.02 | 0.99 |
| Aspirin within 3 days prior to randomization (Y) | 20.51% | 0.67 | 19.14% | 0.66 | 0.02 | 0.98 | 19.29% | 0.66 | 19.14% | 0.66 | 0 | 1 | 20.22% | 0.67 | 20.23% | 0.67 | 0 | 1 |
| Atrial fibrillation (N) | 79.27% | 0.94 | 75.6% | 0.93 | 0.04 | 0.97 | 76.89% | 0.94 | 75.6% | 0.93 | 0.01 | 0.99 | 78.66% | 0.94 | 79.31% | 0.94 | 0.01 | 0.99 |
| Atrial fibrillation (unknown) | 5.11% | 0.48 | 4.88% | 0.47 | 0 | 1 | 2.26% | 0.39 | 4.88% | 0.47 | 0.07 | 0.95 | 5.06% | 0.47 | 4.25% | 0.45 | 0.02 | 0.99 |
| Atrial fibrillation (Y) | 15.62% | 0.63 | 19.52% | 0.66 | 0.06 | 0.95 | 20.85% | 0.68 | 19.52% | 0.66 | 0.02 | 0.98 | 16.28% | 0.64 | 16.44% | 0.64 | 0 | 1 |
| Conscious state at randomization (drowsy) | 17.66% | 0.65 | 41.99% | 0.8 | 0.36 | 0.72 | 42.99% | 0.81 | 41.99% | 0.8 | 0.01 | 0.99 | 21.9% | 0.68 | 21.68% | 0.68 | 0 | 1 |
| Conscious state at randomization (fully alert) | 81.25% | 0.95 | 55.51% | 0.86 | 0.28 | 0.78 | 54.88% | 0.86 | 55.51% | 0.86 | 0.01 | 0.99 | 76.78% | 0.94 | 77.15% | 0.94 | 0 | 1 |
| Conscious state at randomization (unconscious) | 1.09% | 0.32 | 2.5% | 0.4 | 0.04 | 0.97 | 2.13% | 0.38 | 2.5% | 0.4 | 0.01 | 0.99 | 1.32% | 0.34 | 1.17% | 0.33 | 0 | 1 |
| CT before randomization (N) | 33.15% | 0.76 | 32.11% | 0.75 | 0.01 | 0.99 | 31.34% | 0.75 | 32.11% | 0.75 | 0.01 | 0.99 | 32.92% | 0.76 | 31.9% | 0.75 | 0.01 | 0.99 |

| | | | | | | | | | | | | | | | | |
|---|---|---|---|---|---|---|---|---|---|---|---|---|---|---|---|---|
| CT before randomization (Y) | 66.85% | 0.9 | 67.89% | 0.91 | 0.01 | 0.99 | 68.66% | 0.91 | 67.89% | 0.91 | 0.01 | 0.99 | 67.08% | 0.9 | 68.1% | 0.91 | 0.01 | 0.99 |
| Face deficit (can't asses) | 1.03% | 0.32 | 2.41% | 0.39 | 0.04 | 0.97 | 2.27% | 0.39 | 2.41% | 0.39 | 0 | 1 | 1.25% | 0.33 | 1.08% | 0.32 | 0.01 | 1 |
| Face deficit (N) | 27.37% | 0.72 | 20.45% | 0.67 | 0.1 | 0.92 | 20.56% | 0.67 | 20.45% | 0.67 | 0 | 1 | 26.12% | 0.71 | 25.51% | 0.71 | 0.01 | 0.99 |
| Face deficit (Y) | 71.6% | 0.92 | 77.14% | 0.94 | 0.06 | 0.95 | 77.16% | 0.94 | 77.14% | 0.94 | 0 | 1 | 72.62% | 0.92 | 73.42% | 0.93 | 0.01 | 0.99 |
| Arm/hand deficit (can't assess) | 0.54% | 0.27 | 1.07% | 0.32 | 0.02 | 0.98 | 0.93% | 0.31 | 1.07% | 0.32 | 0 | 1 | 0.64% | 0.28 | 0.68% | 0.29 | 0 | 1 |
| Arm/hand deficit (N) | 14.27% | 0.61 | 11.1% | 0.58 | 0.05 | 0.96 | 10.74% | 0.57 | 11.1% | 0.58 | 0.01 | 0.99 | 13.69% | 0.61 | 12.61% | 0.6 | 0.02 | 0.99 |
| Arm/hand deficit (Y) | 85.19% | 0.96 | 87.83% | 0.97 | 0.03 | 0.98 | 88.33% | 0.97 | 87.83% | 0.97 | 0.01 | 1 | 85.67% | 0.96 | 86.71% | 0.96 | 0.01 | 0.99 |
| Leg/foot deficit (can't assess) | 1.22% | 0.33 | 1.76% | 0.36 | 0.02 | 0.99 | 1.67% | 0.36 | 1.76% | 0.36 | 0 | 1 | 1.32% | 0.34 | 1.26% | 0.33 | 0 | 1 |
| Leg/foot deficit (N) | 24.04% | 0.7 | 18.96% | 0.66 | 0.07 | 0.94 | 19.14% | 0.66 | 18.96% | 0.66 | 0 | 1 | 23.1% | 0.69 | 21.52% | 0.68 | 0.02 | 0.98 |
| Leg/foot deficit (Y) | 74.74% | 0.93 | 79.29% | 0.94 | 0.05 | 0.96 | 79.19% | 0.94 | 79.29% | 0.94 | 0 | 1 | 75.58% | 0.93 | 77.22% | 0.94 | 0.02 | 0.99 |
| Dysphasia (can't assess) | 2.61% | 0.4 | 4.88% | 0.47 | 0.06 | 0.96 | 4.36% | 0.46 | 4.88% | 0.47 | 0.01 | 0.99 | 3% | 0.42 | 2.92% | 0.41 | 0 | 1 |
| Dysphasia (N) | 54.98% | 0.86 | 44.76% | 0.82 | 0.12 | 0.9 | 44.56% | 0.82 | 44.76% | 0.82 | 0 | 1 | 53.25% | 0.85 | 54.35% | 0.86 | 0.01 | 0.99 |
| Dysphasia (Y) | 42.41% | 0.81 | 50.36% | 0.84 | 0.1 | 0.92 | 51.08% | 0.85 | 50.36% | 0.84 | 0.01 | 0.99 | 43.75% | 0.81 | 42.73% | 0.81 | 0.01 | 0.99 |
| Hemianopia (can't assess) | 18.48% | 0.66 | 28.9% | 0.73 | 0.16 | 0.87 | 29.33% | 0.74 | 28.9% | 0.73 | 0.01 | 1 | 20.26% | 0.67 | 18.93% | 0.66 | 0.02 | 0.98 |
| Hemianopia (N) | 66.23% | 0.9 | 52.02% | 0.85 | 0.16 | 0.87 | 51.8% | 0.85 | 52.02% | 0.85 | 0 | 1 | 63.78% | 0.89 | 65.9% | 0.9 | 0.02 | 0.98 |
| Hemianopia (Y) | 15.29% | 0.63 | 19.08% | 0.66 | 0.06 | 0.95 | 18.87% | 0.66 | 19.08% | 0.66 | 0 | 1 | 15.97% | 0.63 | 15.16% | 0.62 | 0.01 | 0.99 |
| Visuospatial disorder (can't assess) | 16.02% | 0.63 | 26.04% | 0.71 | 0.16 | 0.88 | 26.7% | 0.72 | 26.04% | 0.71 | 0.01 | 0.99 | 17.73% | 0.65 | 16.97% | 0.64 | 0.01 | 0.99 |
| Visuospatial disorder (N) | 68.56% | 0.91 | 53.27% | 0.85 | 0.17 | 0.86 | 52.48% | 0.85 | 53.27% | 0.85 | 0.01 | 0.99 | 65.87% | 0.9 | 66.95% | 0.9 | 0.01 | 0.99 |
| Visuospatial disorder (Y) | 15.42% | 0.63 | 20.68% | 0.67 | 0.08 | 0.93 | 20.82% | 0.68 | 20.68% | 0.67 | 0 | 1 | 16.4% | 0.64 | 16.08% | 0.63 | 0.01 | 1 |
| Brainstem/cerebellar signs (can't assess) | 7.24% | 0.52 | 12.71% | 0.6 | 0.1 | 0.92 | 12.6% | 0.6 | 12.71% | 0.6 | 0 | 1 | 8.14% | 0.53 | 7.34% | 0.52 | 0.02 | 0.99 |
| Brainstem/cerebellar signs (N) | 81.57% | 0.95 | 77.17% | 0.94 | 0.05 | 0.96 | 77.28% | 0.94 | 77.17% | 0.94 | 0 | 1 | 80.87% | 0.95 | 82.49% | 0.95 | 0.02 | 0.99 |

| | | | | | | | | | | | | | | | | |
|---|---|---|---|---|---|---|---|---|---|---|---|---|---|---|---|---|
| Brainstem/cerebellar signs (Y) | 11.19% | 0.58 | 10.12% | 0.56 | 0.02 | 0.98 | 10.12% | 0.56 | 10.12% | 0.56 | 0 | 1 | 10.99% | 0.58 | 10.16% | 0.56 | 0.01 | 0.99 |
| Other deficit (can't assess) | 5.74% | 0.49 | 9.73% | 0.56 | 0.08 | 0.94 | 10.41% | 0.57 | 9.73% | 0.56 | 0.01 | 0.99 | 6.41% | 0.5 | 6.23% | 0.5 | 0 | 1 |
| Other deficit (N) | 88.01% | 0.97 | 83.84% | 0.96 | 0.04 | 0.97 | 83.27% | 0.96 | 83.84% | 0.96 | 0.01 | 1 | 87.31% | 0.97 | 87.98% | 0.97 | 0.01 | 0.99 |
| Other deficit (Y) | 6.25% | 0.5 | 6.43% | 0.5 | 0 | 1 | 6.33% | 0.5 | 6.43% | 0.5 | 0 | 1 | 6.28% | 0.5 | 5.79% | 0.49 | 0.01 | 0.99 |
| Region (East Asia and Pacific) | 6.87% | 0.51 | 6.28% | 0.5 | 0.01 | 0.99 | 6.87% | 0.51 | 6.28% | 0.5 | 0.01 | 0.99 | 6.79% | 0.51 | 7.94% | 0.53 | 0.02 | 0.98 |
| Region (Europe and Central Asia) | 84.4% | 0.96 | 86.07% | 0.96 | 0.02 | 0.99 | 84.4% | 0.96 | 86.07% | 0.96 | 0.02 | 0.99 | 84.68% | 0.96 | 82.78% | 0.95 | 0.02 | 0.98 |
| Region (Latin America and Caribbean) | 3.74% | 0.44 | 2.74% | 0.41 | 0.02 | 0.98 | 3.74% | 0.44 | 2.74% | 0.41 | 0.02 | 0.98 | 3.56% | 0.43 | 4.52% | 0.46 | 0.02 | 0.98 |
| Region (Middle East and North Africa) | 2.05% | 0.38 | 2.11% | 0.38 | 0 | 1 | 2.05% | 0.38 | 2.11% | 0.38 | 0 | 1 | 2.08% | 0.38 | 2.17% | 0.38 | 0 | 1 |
| Region (North America) | 1.37% | 0.34 | 0.83% | 0.3 | 0.02 | 0.99 | 1.37% | 0.34 | 0.83% | 0.3 | 0.02 | 0.99 | 1.28% | 0.34 | 1% | 0.32 | 0.01 | 0.99 |
| Region (South Asia) | 1.25% | 0.33 | 1.4% | 0.34 | 0 | 1 | 1.25% | 0.33 | 1.4% | 0.34 | 0 | 1 | 1.27% | 0.34 | 1.34% | 0.34 | 0 | 1 |
| Region (Sub-Saharan Africa) | 0.31% | 0.24 | 0.57% | 0.27 | 0.01 | 0.99 | 0.31% | 0.24 | 0.57% | 0.27 | 0.01 | 0.99 | 0.34% | 0.24 | 0.25% | 0.22 | 0 | 1 |
| Heparin within 24 hours prior to randomization (N) | 95.79% | 0.99 | 96.9% | 0.99 | 0.01 | 0.99 | 97.39% | 0.99 | 96.9% | 0.99 | 0 | 1 | 95.98% | 0.99 | 96.04% | 0.99 | 0 | 1 |
| Heparin within 24 hours prior to randomization (unknown) | 1.84% | 0.37 | 1.46% | 0.35 | 0.01 | 0.99 | 0.45% | 0.26 | 1.46% | 0.35 | 0.04 | 0.97 | 1.76% | 0.36 | 1.37% | 0.34 | 0.01 | 0.99 |
| Heparin within 24 hours prior to randomization (Y) | 2.38% | 0.39 | 1.64% | 0.36 | 0.02 | 0.98 | 2.17% | 0.38 | 1.64% | 0.36 | 0.01 | 0.99 | 2.26% | 0.39 | 2.59% | 0.4 | 0.01 | 0.99 |
| Symptoms noted on waking (N) | 70.63% | 0.92 | 71.34% | 0.92 | 0.01 | 0.99 | 71.4% | 0.92 | 71.34% | 0.92 | 0 | 1 | 70.7% | 0.92 | 71.2% | 0.92 | 0 | 1 |
| Symptoms noted on waking (Y) | 29.37% | 0.74 | 28.66% | 0.73 | 0.01 | 0.99 | 28.6% | 0.73 | 28.66% | 0.73 | 0 | 1 | 29.3% | 0.74 | 28.8% | 0.73 | 0.01 | 0.99 |
| Infarct visible on CT (N) | 67.59% | 0.91 | 64.17% | 0.89 | 0.04 | 0.97 | 63.57% | 0.89 | 64.17% | 0.89 | 0.01 | 0.99 | 66.86% | 0.9 | 64.87% | 0.9 | 0.02 | 0.98 |
| Infarct visible on CT (Y) | 32.41% | 0.75 | 35.83% | 0.77 | 0.05 | 0.96 | 36.43% | 0.78 | 35.83% | 0.77 | 0.01 | 0.99 | 33.14% | 0.76 | 35.13% | 0.77 | 0.03 | 0.98 |
| Trial heparin allocated (high dose, pilot phase) | 1.29% | 0.34 | 1.13% | 0.33 | 0 | 1 | 0.46% | 0.26 | 1.13% | 0.33 | 0.03 | 0.98 | 1.27% | 0.34 | 1.13% | 0.33 | 0 | 1 |
| Trial heparin allocated (low dose) | 25% | 0.71 | 25.18% | 0.71 | 0 | 1 | 24.18% | 0.7 | 25.18% | 0.71 | 0.01 | 0.99 | 24.99% | 0.71 | 23.49% | 0.7 | 0.02 | 0.98 |
| Trial heparin allocated (high dose) | 23.65% | 0.7 | 23.99% | 0.7 | 0 | 1 | 25.61% | 0.71 | 23.99% | 0.7 | 0.02 | 0.98 | 23.73% | 0.7 | 23.25% | 0.69 | 0.01 | 0.99 |

| | | | | | | | | | | | | | | | | | |
|---|---|---|---|---|---|---|---|---|---|---|---|---|---|---|---|---|---|
| Trial heparin allocated (N) | 50.06% | 0.84 | 49.7% | 0.84 | 0 | 1 | 49.75% | 0.84 | 49.7% | 0.84 | 0 | 1 | 50.01% | 0.84 | 52.13% | 0.85 | 0.03 | 0.98 |
| Sex (F) | 45.87% | 0.82 | 49.32% | 0.84 | 0.04 | 0.97 | 51.58% | 0.85 | 49.32% | 0.84 | 0.03 | 0.98 | 46.86% | 0.83 | 49.74% | 0.84 | 0.04 | 0.97 |
| Sex (M) | 54.13% | 0.86 | 50.68% | 0.84 | 0.04 | 0.97 | 48.42% | 0.83 | 50.68% | 0.84 | 0.03 | 0.98 | 53.14% | 0.85 | 50.26% | 0.84 | 0.03 | 0.97 |
| Stroke subtype (lacunar) | 25.65% | 0.71 | 15.89% | 0.63 | 0.14 | 0.89 | 15.51% | 0.63 | 15.89% | 0.63 | 0.01 | 1 | 23.99% | 0.7 | 26.9% | 0.72 | 0.04 | 0.97 |
| Stroke subtype (other) | 0.31% | 0.24 | 0.21% | 0.21 | 0 | 1 | 0.26% | 0.23 | 0.21% | 0.21 | 0 | 1 | 0.3% | 0.23 | 0.37% | 0.25 | 0 | 1 |
| Stroke subtype (partial anterior circulation) | 40.51% | 0.8 | 40.03% | 0.8 | 0.01 | 1 | 40.3% | 0.8 | 40.03% | 0.8 | 0 | 1 | 40.4% | 0.8 | 39.81% | 0.79 | 0.01 | 0.99 |
| Stroke subtype (posterior circulation) | 11.7% | 0.58 | 10.33% | 0.57 | 0.02 | 0.98 | 10.29% | 0.57 | 10.33% | 0.57 | 0 | 1 | 11.45% | 0.58 | 10.56% | 0.57 | 0.02 | 0.99 |
| Stroke subtype (total anterior circulation) | 21.83% | 0.68 | 33.54% | 0.76 | 0.17 | 0.86 | 33.63% | 0.76 | 33.54% | 0.76 | 0 | 1 | 23.86% | 0.7 | 22.36% | 0.69 | 0.02 | 0.98 |

**Appendix References**

World Bank (2021), World Bank Country and Lending Groups. https://datahelpdesk.worldbank.org/knowledgebase/articles/906519-world-bank-country-and-lending-groups [accessed February 17, 2022].